\documentclass[aps,pra,twocolumn]{revtex4}
\usepackage{amsmath}
\usepackage{amscd}
\usepackage{graphicx}
\usepackage{subfigure}
\usepackage{appendix}

\begin{document}
\title[Short Title]{Fast quantum state engineering via universal SU(2) transformation  }

\author{Bi-Hua Huang$^{1,2}$}
\author{Yi-Hao Kang$^{1,2}$}
\author{Ye-Hong Chen$^{1,2}$}
\author{Qi-Cheng Wu$^{1,2}$}
\author{Jie Song$^{3}$}
\author{Yan Xia$^{1,2,}$\footnote{E-mail: xia-208@163.com}}

\affiliation{$^{1}$Department of Physics, Fuzhou University, Fuzhou 350116, China\\
             $^{2}$Fujian Key Laboratory of Quantum Information and Quantum Optics (Fuzhou University), Fuzhou 350116, China\\
             $^{3}$Department of Physics, Harbin Institute of Technology, Harbin 150001, China}

\begin{abstract}
We introduce a simple yet versatile protocol to inverse engineer the time-dependent Hamiltonian in two- and three level systems. In the protocol, by utilizing a universal SU(2) transformation, a given speedup goal can be obtained with large freedom to select the control parameters.
As an illustration example, the protocol is applied to perform population transfer between nitrogen-vacancy (NV) centers in diamond. Numerical
 simulation shows that the speed of the present protocol is fast compared with that
 of the adiabatic process. Moreover, the protocol is also
 tolerant to decoherence and
experimental parameter fluctuations. Therefore, the protocol may be useful
for designing an experimental feasible Hamiltonian to engineer a quantum system.
\end{abstract}

\maketitle

\section{Introduction}

Controlling the system dynamics with time-dependent external fields
is of great interest in recent decades. In an effort to develop
control tools to engineer the dynamics of the system, methods which
allow one to design fast and robust quantum control pulses have been put
forward, such as optimal control theory
\cite{Sugny0877,Peirce8737,GlaserEPJD1569} and composite pulse sequences \cite{Daems13111,VitanovPRL13110,VitanovPRA1183}, etc..
Recently, by designing nonadiabatic shortcuts to speed up the quantum adiabatic process, a new method named ``shortcuts to adiabaticity" (STA) \cite{DemirplakJPCA03107,DemirplakJCP08129107,MugaAAMOP1362,MugaJPB0942,ChenXPRL10104,AdelSR1202,ChenXPRA1183,BerryJPA0942,AdelPRA1184,MasudaPRSA09466,MasudaPRA1184,TorronteguiPRA1286,ChenXPRL10105,AdelPRL13111} has become an interesting subject quietly into the researchers' vision and attracted an increasing amount of theoretical and
experimental research interest \cite{SongNJP1618, ChenZSR1606,
ChenYHSR1605, ChenYHOC16380, ShanWJQIP1615, SarandySR1505,
SarandyPRA1693, LiangPRA1591, SarandyFI1603, AdelNJP1618,
AdelPRX1404, AdelPRL12109, AdelNC1607, DuNC1607, ZhangJPRL13110,
TorronteguiPRA1183, MugaJPB1043, TorronteguiPRA1285, ChenYHPRA1591,
ChenXPRA1082, SchaffNJP1113, SchaffPRA1082, SchaffEPL1193,
AdelEPL1196, TorronteguiNJP1214,
LuMPRA1489,AdelPRL17118,KangPRA1694,ChenYHPRA1489,ZhongSR16,BasonNP1208,Ruschhaupt1214}.

In STA, the Hamiltonian is designed to drive the system following a
nonadiabatic path that reproduces the same final state as the
adiabatic process does in a relatively short time. By means of
different methods, e.g., transitionless quantum driving (TQD)
\cite{DemirplakJPCA03107,DemirplakJCP08129107,BerryJPA0942,ChenXPRL10105,AdelPRL13111},
inverse engineering based on Lewis-Riesenfeld (LR)
\cite{LewisJMP6910} invariants \cite{MugaJPB0942,ChenXPRA1183,
ChenXPRL10104}, and so on
\cite{AdelPRA1184,AdelSR1202,MasudaPRSA09466,MasudaPRA1184,TorronteguiPRA1286},
a variety of protocols has been developed to engineer STA. Among
them, the invariant-based protocol is based on first designing an LR
invariant, then constructing from the invariant a transient driving
Hamiltonian. But for most systems, the invariants are unknown or
hard to be solved. As for TQD, the shortcut is constructed by
adding to the original time-dependent Hamiltonian a specific
designed ``counterdiabatic"~(CD) term, which requires knowledge of
the spectral properties of the origin Hamiltonian at all times. This
constraint has limited the range of applicability of the protocol.
And, sometimes the constructed CD term is difficult or impossible to
realize in the laboratory \cite{AdelPRL12109,MugaAAMOP1362}. To
overcome the difficulty, several protocols
\cite{BaksicPRL16116,OpatrnyNJP1416,OpatrnyPRA1490,TorronteguiPRA1489,TorosovPRA1489,ChenYHPRA1693,KangYHSR1606,IbanezPRL12109,IbanezPRA1387,SongXKPRA1693,MartinezPRA1489,HuangBHLPL16,LiPRA1694}
have been put forward. For example, Ib\'{a}\~{n}ez \textit{et al.}
\cite{IbanezPRL12109, IbanezPRA1387} have proposed to use the iterative
picture to construct shortcuts. Baksic \textit{et al.}
\cite{BaksicPRL16116} have advised speeding up the TQD by employing
a set of dressed states. Chen \textit{et al.} \cite{ChenYHPRA1693}
have suggested constructing STA with alternative CD terms. Those
protocols provide us a lot of options to achieve fast quantum state
engineering and each one has its own characteristics. Therefore, it
is meaningful to delve deep to enrich the methods used in STA for
various application situations.

In this paper, inspired by past theories, we propose a different protocol
for quantum state engineering in two- and three level systems. The basic strategy of the protocol is to use a universal SU(2) transformation to design the evolution path of the system.
To make the basic idea clearly, we illustrate the construction algorithm
by applying it to fulfill the population transfer process in distant nitrogen-vacancy~(NV) centers in diamond, which is coupled to the quantized whispering-gallery mode~(WGM) of a fused-silica high-$Q$ microsphere cavity.
This NV-WGM composite system takes advantage of both sides of NV centers and WGM microcavity, i.e., WGM microcavity possesses ultrahigh $Q$ factors and small mode volume, which enables strong temporal and spatial confinement of photons \cite{ThomasNL0606,GorodetskyOL9621,VernooyOL9823,XiaoPRA1285}. Meanwhile, the NV center in diamond is viewed as an excellent candidate for quantum information processing, owing to its sufficiently long electronic spin lifetime and the possibility of coherent manipulation at room temperature \cite{ChildressSC06314,ToganNature10466}. Apart from the advantages of employing the NV-WGM system, our protocol also holds the following advantages: (1) By designing the moving state of the system with different control parameters, it provides
a variety of optional paths for quantum evolution between
an initial state and a final state, while the adiabatic condition,
which limits the evolution speed, could be abandoned.
(2) The present protocol not only shares the advantage with the LR invariant-based protocol, but also renders more freedom for choosing parameters to design pulses, which makes the protocol more flexible.
(3) According to the numerical simulation, it also holds high speed and good robustness against parameter deviations and dissipations. Therefore, the protocol may enrich the shortcut methods and provide us an alternative choice for the fast implementation of quantum information processing.

The paper is organized as follows. In Sec.~\ref{section:II}, we will introduce the constructing method for inverse engineering a Hamiltonian based on universal SU(2) transformation and give a general framework of the protocol. In Sec.~\ref{section:III},
 we will illustrate the shortcut protocol by applying it to implement quantum population transfer with the NV centers coupled to a WGM microcavity in detail, and check out the robustness of the present protocol by numerical
simulations. Conclusions will be present in Sec.~\ref{section:IV}.

\section{GENERAL SHORTCUTS FOR SU(2) SYMMETRIC SYSTEMS VIA ROTATION TRANSFORMATION}\label{section:II}

It is known, for any two-level quantum system, the Hamiltonian can be expressed as $(\hbar=1)$,
  \begin{eqnarray}\label{eqb1}
     H_0(t)= g_x(t)\sigma_x+g_y(t)\sigma_y+g_z(t)\sigma_z,
  \end{eqnarray}
where $g_k(t)~(k=x,y,z)$ are arbitrary real functions of
time, and $\sigma_x,\sigma_y,\sigma_z$ are Pauli operators.

 Let us consider the time-dependent Schr\"{o}dinger equation describing a quantum system is given by
  \begin{eqnarray}\label{eqb1}
  i\partial_t|\psi(t)\rangle=H_0(t)|\psi(t)\rangle,
  \end{eqnarray}
where $|\psi(t)\rangle$ is the wave function and can be expressed by the evolution operator $\textrm{U}_O(t)$ as $|\psi(t)\rangle=\textrm{U}_O(t)|\psi(0)\rangle$. In order to inverse design feasible Hamiltonians (and shortcuts) $H_0(t)$ that give the desired dynamics, we perform a picture transformation defined by $R^{\dag}$ as $|\Psi(t)\rangle=R^{\dag}|\psi(t)\rangle$, where $R$ is the most general unitary transformation of a two-level system \cite{VitanovPRA1285}:
  \begin{eqnarray}\label{eqb2}
    R(\theta,\xi,\eta)&=&
                    \left(
                     \begin{array}{cccc}
                       e^{i\xi}\cos\theta  &  -e^{-i\eta}\sin\theta  \\
                       e^{i\eta}\sin\theta  &  e^{-i\xi}\cos\theta  \\
                    \end{array}
                    \right)
                    \cr\cr\cr
                    &=&e^{\frac{i(\xi-\eta)}{2}\sigma_{z}}e^{-i\theta\sigma_{y}}e^{\frac{i(\xi+\eta)}{2}\sigma_{z}}.
  \end{eqnarray}
  According to picture transformation, we have
  \begin{eqnarray}\label{eqb2}
  i\partial_t|\Psi(t)\rangle=H(t)|\Psi(t)\rangle,~~
  \cr\cr
  H(t)=R^{\dag}H_0(t)R+i\partial_t{R^{\dag}}R.
  \end{eqnarray}
  Therefore, if we could properly choose the parameters $(\xi,\eta,\theta)$ so that $H(t)$ can be designed as $H(t)=f_k(t)\sigma_k~(k=x,y,z)$, the evolution operator in the picture $R$ will be $\textrm{U}_R(t)=e^{-i\int^{t}_0 H(t^{\prime})dt^{\prime}}=e^{-i\delta_k\sigma_k}$, $\delta_k=\int^{t}_0 f_k(t^{\prime})dt^{\prime}$.
    Then back to the Schr\"{o}dinger picture, we will obtain
    \begin{eqnarray}\label{eqb2}
\textrm{U}_O(t)=R(t)\textrm{U}_R(t)R^{\dag}(0).
 \end{eqnarray}
     Particularly, in the case of $R(0)=1$, we have
 $|\Psi(0)\rangle=|\psi(0)\rangle$ and $\textrm{U}_O(t)=R(t)\textrm{U}_R(t)$.
  Since $\textrm{U}_O(t)$ can be expressed as $\textrm{U}_O(t)=\sum_n|\phi_n(t)\rangle\langle n|$, where $\{|n\rangle\}$ and $\{|\phi_n(t)\rangle\}$ are respectively the bare states and the moving states of the system, we could obtain the moving states $\{|\phi_n(t)\rangle\}$ as ``riding bus'' to drive the system along the desired dynamics.

    In other words, our idea can be summarized as follows: (1) According to Eq.~(5), by designing $R(t)$ and $\textrm{U}_R(t)$, the evolution operator of the system [$\textrm{U}_O(t)$] and the pulses $g_k(t)$
    [both are parameterized with time-dependent parameters
($\xi,\theta,\eta$)] are designed.
    (2) $\textrm{U}_O(t)$ is rewritten to obtain the moving states of the system.
    (3) Based on the desired dynamics and the moving state of the system, boundary conditions of the parameters $(\xi, \eta, \theta)$ are specified.
    (4) With the boundary conditions, the functions of the parameters $(\xi, \eta, \theta)$ are properly designed, accordingly, the pulses $g_k(t)$ are determined, that is, the inverse designation of Hamiltonian $H_0(t)$ is achieved.
    In the following, we will elaborate the details.

  For brevity, we set $\xi-\eta=\alpha$ and $\xi+\eta=\beta$, it is not difficult to obtain the explicit forms of $R^{\dag}H_0(t)R$ and $\partial_t{R^{\dag}}R$:
\begin{eqnarray}\label{eqb3}
  R^{\dag}H_0(t)R&=&\sigma_{x}[g_{x}\cos\alpha\cos\beta\cos2\theta-g_{x}\sin\alpha\sin\beta
                 \cr
                 &-&g_{y}\sin\alpha\cos\beta\cos2\theta-g_{y}\cos\alpha\sin\beta
                 \cr
                 &-&g_{z}\cos\beta\sin2\theta]
                 \cr
               &+&\sigma_{y}[g_{x}\cos\alpha\sin\beta\cos2\theta+g_{x}\sin\alpha\cos\beta
               \cr
               &-&g_{y}\sin\alpha\sin\beta\cos2\theta+g_{y}\cos\alpha\cos\beta
               \cr
               &-&g_{z}\sin\beta\sin2\theta]
               \cr
               &+&\sigma_{z}[g_{x}\cos\alpha\sin2\theta
               \cr
               &-&g_{y}\sin\alpha\sin2\theta
               \cr
               &+&g_{z}\cos2\theta],
\end{eqnarray}
\begin{eqnarray}\label{eqb4}
  \partial_t{R^{\dag}}R&=&\sigma_{x}[\dot{\theta}\sin\beta-\dot{\alpha}\cos\beta\sin\theta\cos\theta]  \cr
                       &+&\sigma_{y}[-\dot{\theta}\cos\beta-\dot{\alpha}\sin\beta\sin\theta\cos\theta]  \cr
                       &+&\sigma_{z}[\frac{\dot{\beta}}{2}+\frac{\dot{\alpha}}{2}\cos2\theta].
\end{eqnarray}
Next, we will take $\beta=0$ and $\alpha=0$ to illustrate the
detailed constructing procedure, respectively. For simplicity, we consider $g_y=0$.

Case I: $\beta=0,~g_y=0$. Hamiltonian $H(t)$ can be simplified as
 \begin{eqnarray}\label{eqb5}
     H(t)&=&R^{\dag}H_0(t)R+i\partial_t{R^{\dag}}R \cr
     &=&\sigma_{x}[g_{x}\cos\alpha\cos2\theta-g_{z}\sin2\theta-\dot{\alpha}\sin\theta\cos\theta] \cr
           &+&\sigma_{y}[g_{x}\sin\alpha-\dot{\theta}]
           \cr
           &+&\sigma_{z}[g_{x}\cos\alpha\sin2\theta+g_{z}\cos2\theta+\frac{\dot{\alpha}}{2}\cos2\theta].
  \end{eqnarray}
 By setting $H(t)$ equal to either one of $f_k(t)\sigma_k~(k=x,y,z)$, we can obtain the explicit forms of $g_x,~g_z$. For example, if we choose  $H(t)=f_z(t)\sigma_z$, which implies that the coefficients of the terms $\sigma_x$ and $\sigma_y$ in Eq.~(8) are equal to zero, then we obtain
\begin{eqnarray}\label{eqb6}
     g_x=\frac{\dot{\theta}}{\sin\alpha},~~g_z=\dot{\theta}\cot\alpha\cot2\theta-\frac{\dot{\alpha}}{2},
  \end{eqnarray}
and
\begin{eqnarray}\label{eqb7}
     f_z(t)=\frac{\dot{\theta}\cot\alpha}{\sin2\theta}.
  \end{eqnarray}
 Setting $\delta_z=\int^{t}_0\frac{2\dot{\theta}\cot\alpha}{\sin2\theta} dt^{
\prime}$ and $R(0)=1$, according to Eqs.~(3) and (5), the evolution
operator $\textrm{U}_O(t)$ is
\begin{eqnarray}\label{eqb8}
     \textrm{U}_O(t)&=&e^{\frac{i\alpha}{2}\sigma_z}e^{-i\theta\sigma_y}e^{-\frac{i\delta_z}{2}\sigma_z}
            \cr\cr
            &=&\left(
             \begin{array}{cccc}
               e^{\frac{i\alpha-i\delta_z}{2}}\cos\theta & ~~-e^{\frac{i\alpha+i\delta_z}{2}}\sin\theta  \\
               e^{\frac{-i\alpha-i\delta_z}{2}}\sin\theta &  ~~e^{\frac{-i\alpha+i\delta_z}{2}}\cos\theta \\
             \end{array}
            \right).
  \end{eqnarray}
We suppose $|1\rangle=(1,0)^T$ and $|2\rangle=(0,1)^T$. Then,
rewriting $\textrm{U}_O(t)=\sum_{n=1}^{2}|\phi_n(t)\rangle\langle
n|$, we can obtain the moving states of the system, i.e.,
$|\phi_1(t)\rangle=(e^{\frac{i\alpha-i\delta_z}{2}}\cos\theta,e^{\frac{-i\alpha-i\delta_z}{2}}\sin\theta)^{T}$,
$|\phi_2(t)\rangle=(-e^{\frac{i\alpha+i\delta_z}{2}}\sin\theta,e^{\frac{-i\alpha+i\delta_z}{2}}\cos\theta)^{T}$. Supposing
the desired dynamics is to achieve a population inverse, from state $|1\rangle$ to state $|2\rangle$, we select
$|\phi_1(t)\rangle$ as the moving state to perform the operation.
 Neglecting the global phase $e^{\frac{i\alpha-i\delta_z}{2}}$ for
$|\phi_1(t)\rangle$, the moving state becomes $(\cos\theta,e^{-i\alpha}\sin\theta)^{T}$.
Thus, the corresponding boundary conditions can be specified
as
\begin{eqnarray}\label{eqb9}
     \theta(0)=0,~~\theta(T)=\frac{\pi}{2},~~\alpha(0)=\alpha(T)=\frac{\pi}{2},
  \end{eqnarray}
 where $T$ is the total evolution time. For the designed pulses could be smoothly switched on and switched off, we choose $\theta$ and $\alpha$ as follows
 \begin{eqnarray}\label{eqb10}
     \theta&=&\frac{\pi}{2}(\frac{t}{T})^2[10-20\frac{t}{T}+15(\frac{t}{T})^2-4(\frac{t}{T})^3],~~~
     \cr
     \dot{\theta}&=&\frac{10\pi}{T}\frac{t}{T}(1-\frac{t}{T})^3,
     \cr
     \alpha&=&\frac{t^2}{T^2}[\frac{1}{2}-\frac{t}{T}+\frac{1}{2}(\frac{t}{T})^2]+\frac{\pi}{2},~~~
     \cr
     \dot{\alpha}&=&\frac{t}{T^2}(1-\frac{t}{T})(1-2\frac{t}{T}).
  \end{eqnarray}
  In this way, according to Eq.~(9), $g_{x}$, $g_{z}$ are determined, namely, we have inversely constructed a shortcut Hamiltonian $H_0(t)$ to realize the population inversion, as seen in Fig.~1.

  It is worth pointing out that, in fact, Eq.~(9) is also the pulses which are designed by the LR invariant method. While, here, it is one of our designing ``recipes''.
  In our protocol, we can choose other parameters to construct the shortcuts, too, i.e., choosing $H(t)=f_x(t)\sigma_x$ or $H(t)=f_y(t)\sigma_y$ or following the constructing procedure in case II.
\begin{figure}
 \scalebox{0.35}{\includegraphics{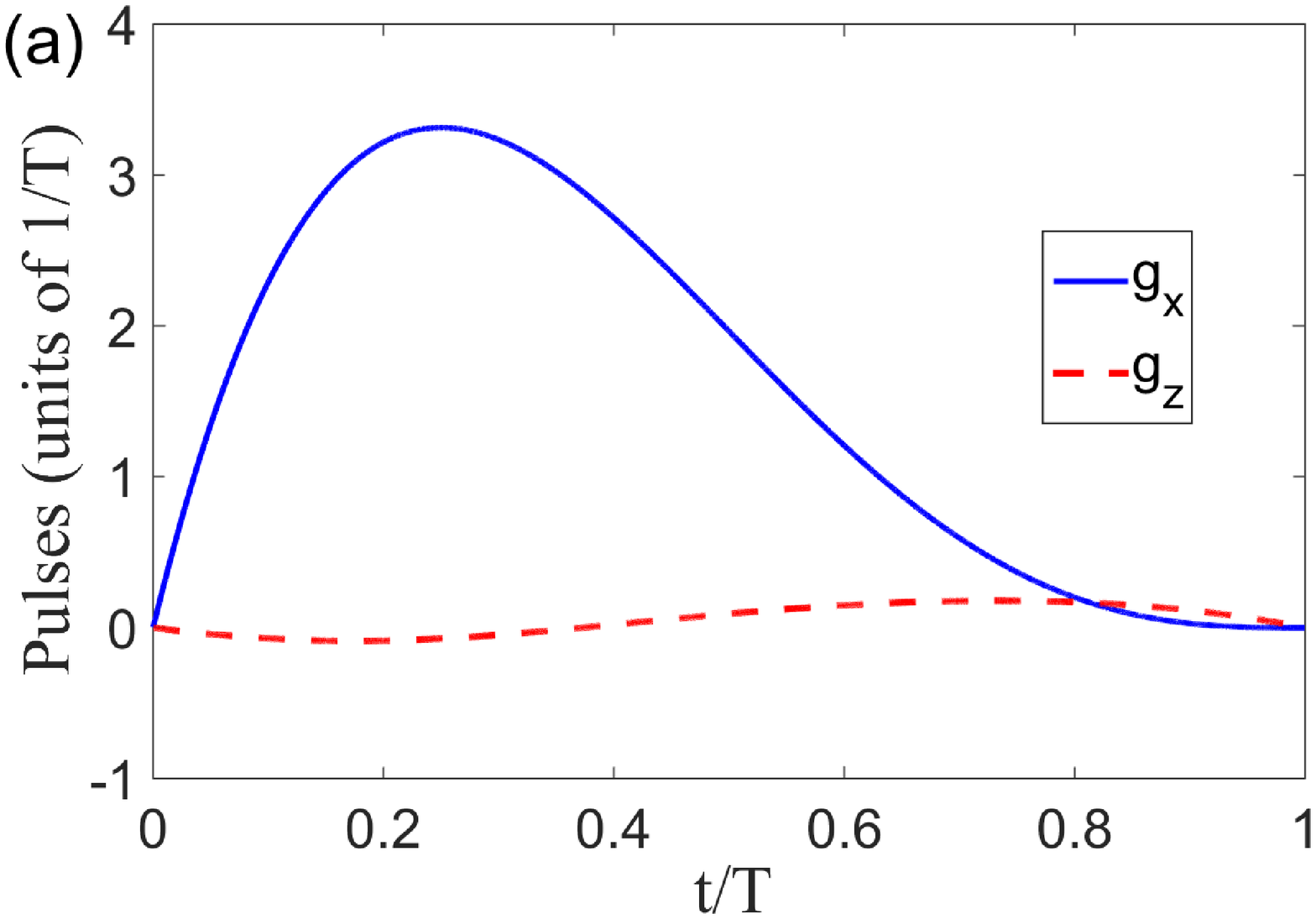}}
 \scalebox{0.35}{\includegraphics{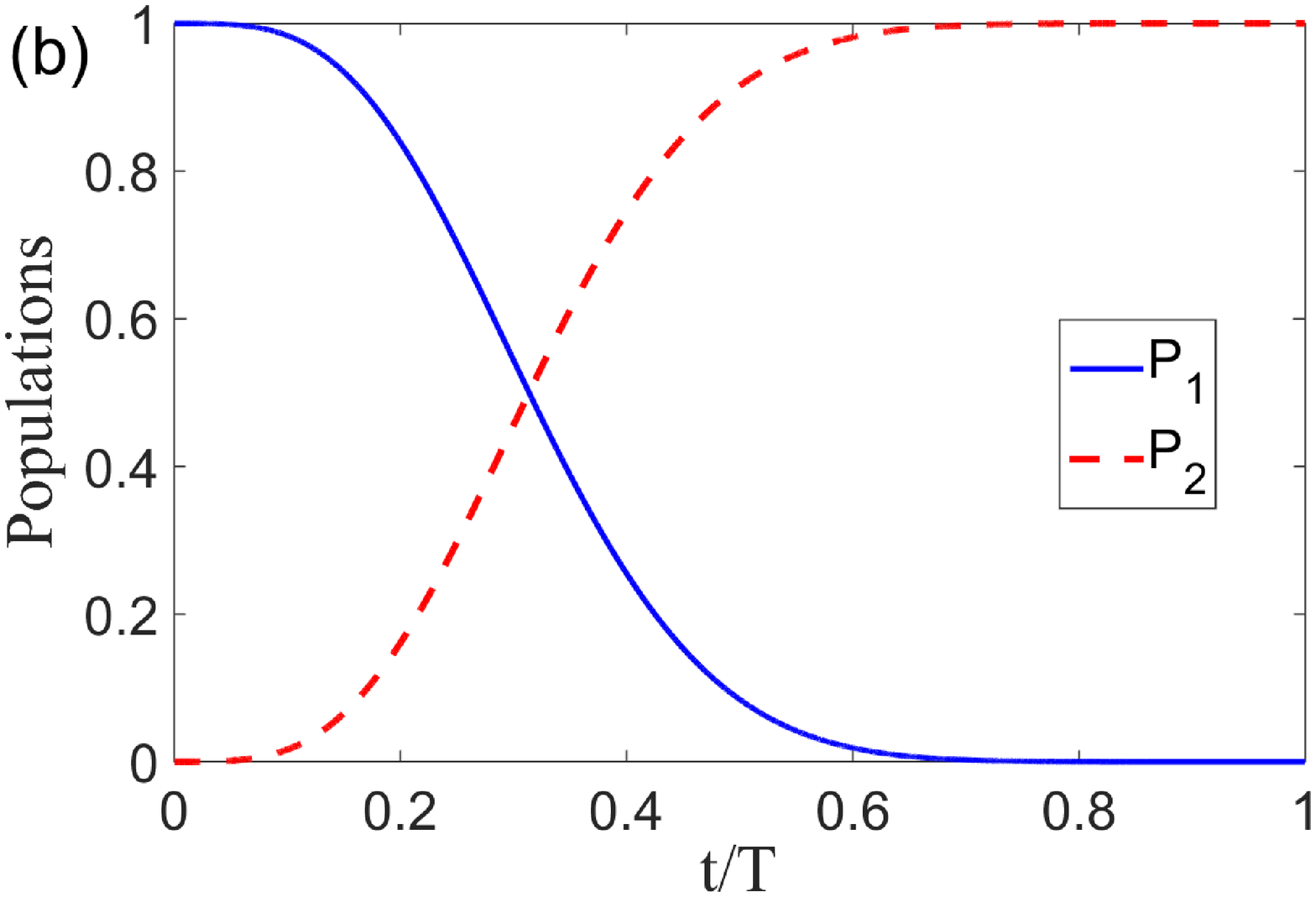}}
 \caption{
         (a) Dependence on $t/T$ of the pulses $g_x, g_z$ for case I.
         (b) Time-evolution for states $|1\rangle$ ($P_1$) and $|2\rangle$ ($P_2$).
         }
 \label{fig1}
\end{figure}

Case II: $\alpha=0,~g_y=0$.
By a similar derivation process as shown in case I, if the Hamiltonian $H(t)$ is chosen as $H(t)=f_x(t)\sigma_x$, we will obtain
\begin{eqnarray}\label{eqb11}
     g_x&=&\dot{\theta}\cos2\theta\cot\beta-\frac{\dot{\beta}}{2}\sin2\theta,
     \cr
     g_z&=&-\dot{\theta}\sin2\theta\cot\beta-\frac{\dot{\beta}}{2}\cos2\theta.
  \end{eqnarray}
Consider that
$R(0)=1/\sqrt{2}(|1\rangle\langle1|+|2\rangle\langle2|+|1\rangle\langle2|-|2\rangle\langle1|)$,
the evolution operator is
\begin{widetext}
\begin{eqnarray}\label{eqb12}
     \textrm{U}_O(t)= \frac{1}{\sqrt{2}}\left(
             \begin{array}{cccc}
               \cos\theta e^{i\frac{\beta-2\delta_x}{2}}-\sin\theta e^{i\frac{-\beta-2\delta_x}{2}} ~~& -\cos\theta e^{i\frac{\beta+2\delta_x}{2}}-\sin\theta e^{i\frac{-\beta+2\delta_x}{2}} \\
               \sin\theta e^{i\frac{\beta-2\delta_x}{2}}+\cos\theta e^{i\frac{-\beta-2\delta_x}{2}}  ~~&  -\sin\theta e^{i\frac{\beta+2\delta_x}{2}}+\cos\theta e^{i\frac{-\beta+2\delta_x}{2}}  \\
             \end{array}
            \right),
  \end{eqnarray}
  \end{widetext}
where $\delta_x=\int^{t}_0
f_x(t^{\prime})dt^{\prime},~f_x(t^{\prime})=\frac{\dot{\theta}}{\sin\beta}$.
The same as that in case I, considering population inversion
$|1\rangle\rightarrow|2\rangle$, the system would evolve through
a moving state $\textrm{U}_O|1\rangle=\frac{1}{\sqrt{2}}(\cos\theta
e^{i\frac{\beta-2\delta_x}{2}}-\sin\theta
e^{i\frac{-\beta-2\delta_x}{2}},~\sin\theta
e^{i\frac{\beta-2\delta_x}{2}}+\cos\theta
e^{i\frac{-\beta-2\delta_x}{2}})^T$. Neglecting a global phase
$e^{i\frac{\beta-2\delta_x}{2}}$, the moving state could be described
as
$\frac{1}{\sqrt{2}}(\cos\theta-e^{-i\beta}\sin\theta,~\sin\theta+e^{-i\beta}\cos\theta)^T$.
Then, for population inverse process, the boundary conditions are:
$\beta(0)=\beta(T)=0$ and
$\theta(0)=-\frac{\pi}{4},~\theta(T)=\frac{\pi}{4}$. By
appropriately designing $\beta$ and $\theta$, another shortcut will
be constructed too. For example, the corresponding $\beta$ and
$\theta$ can be set as
\begin{eqnarray}\label{eqb13}
     \beta&=&A_1(\frac{t}{T})^2[\frac{1}{2}-\frac{t}{T}+\frac{1}{2}(\frac{t}{T})^2],~~~
     \cr
     \dot{\beta}&=&A_1\frac{t}{T^2}(1-\frac{t}{T})(1-2\frac{t}{T}),
     \cr
     \theta&=&\pi(\frac{t}{T})^4[\frac{35}{2}-42\frac{t}{T}+35(\frac{t}{T})^2-10(\frac{t}{T})^3]-\frac{\pi}{4},~~~
     \cr
     \dot{\theta}&=&70\pi(\frac{t}{T})^3(1-\frac{t}{T})^3,
  \end{eqnarray}
where $A_1$ is a tunable time-independent parameter. To avoid the
singularity of the expression and optimize the amplitude for each
pulse, we select $0<A_1<30\pi$. In Fig.~2, we plot the corresponding
pulses and populations with $A_1=8\pi$.
\begin{figure}
 \scalebox{0.35}{\includegraphics{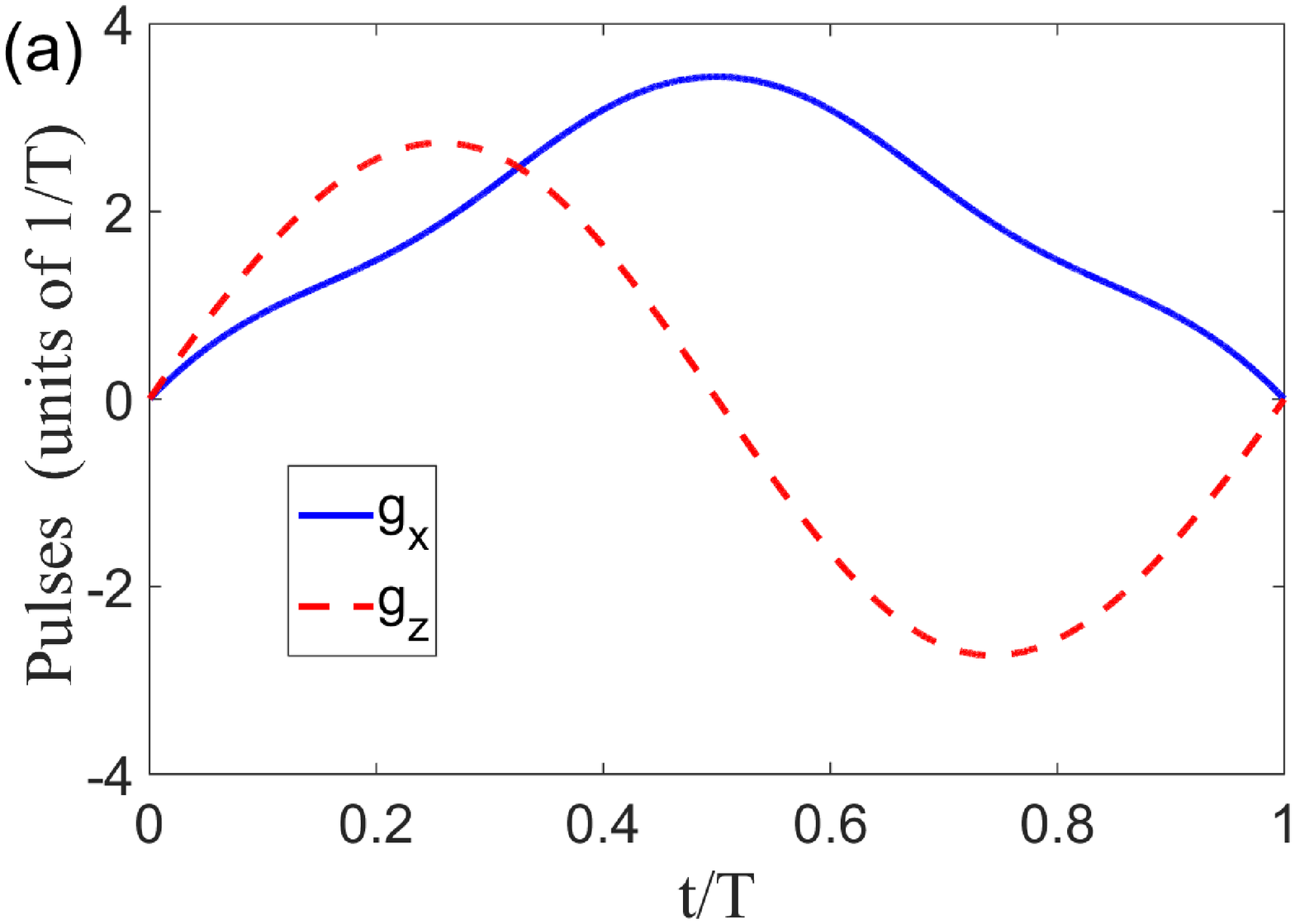}}
 \scalebox{0.35}{\includegraphics{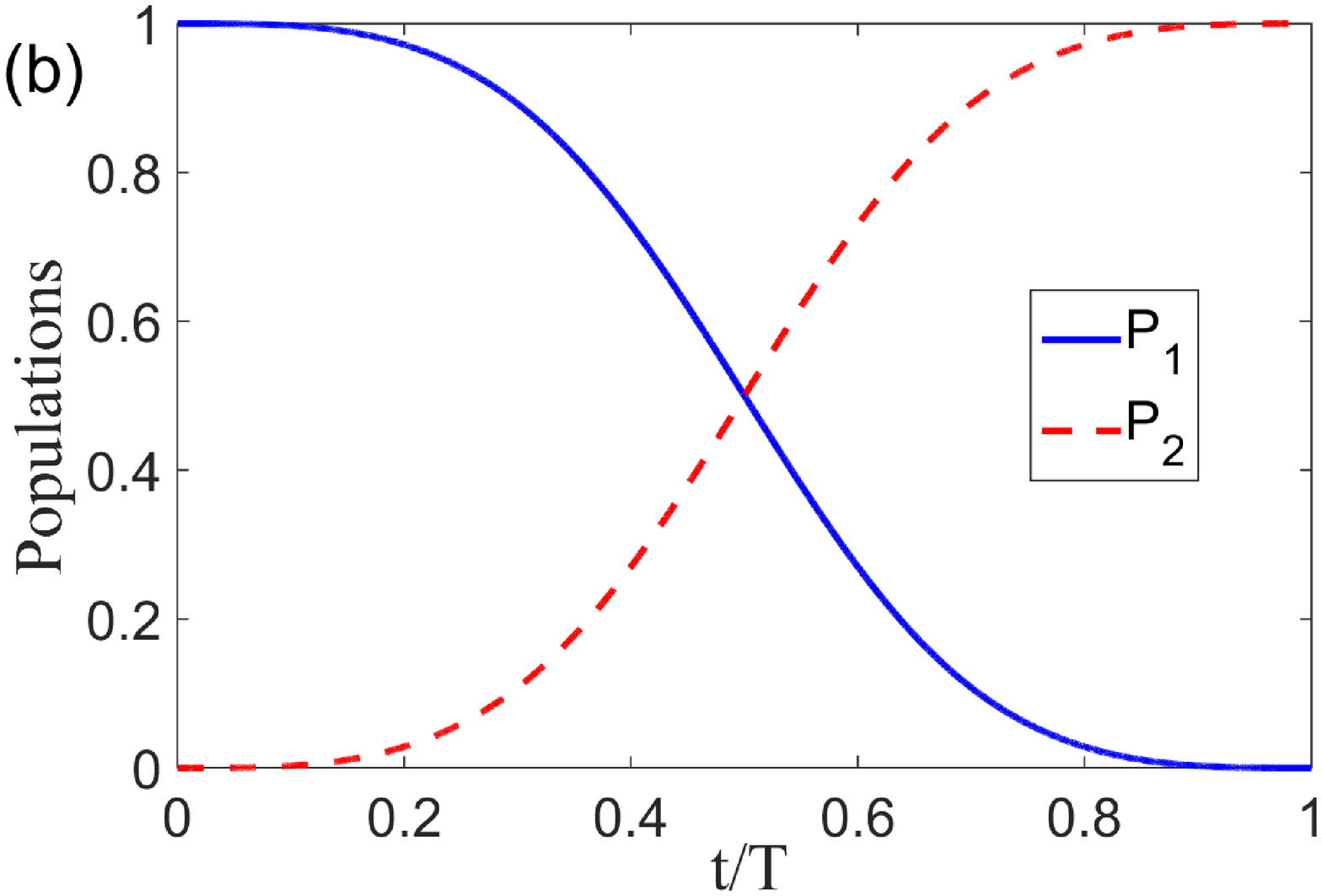}}
 \caption{
         (a) Dependence on $t/T$ of the pulses $g_x, g_z$ for case II with $A_1=8\pi$.
         (b) Time-evolution for states $|1\rangle$ ($P_1$) and $|2\rangle$ ($P_2$).
         }
 \label{fig2}
\end{figure}

Note that the above protocol can be extended to three-level systems with SU(2) symmetry. A meet SU(2) symmetrical Hamiltonian for three-level system can be expressed as
\begin{eqnarray}\label{eqb14}
     H_1^{\prime}(t)= \left(
             \begin{array}{cccc}
               0 & \Omega_x(t) &  0 \\
               \Omega_x(t) &  0 & \Omega_y(t) \\
               0 & \Omega_y(t) &  0 \\
             \end{array}
            \right),
  \end{eqnarray}
where $\Omega_x(t)$, $\Omega_y(t)$ are arbitrary real functions of
time. As Ref.~\cite{HuangBHLPL16} has pointed out, by introducing a unitary matrix
\begin{eqnarray}\label{eqb15}
           V=\frac{1}{\sqrt[]{2}}
             \left(
             \begin{array}{cccc}
               1 & 0 &  1 \\
               0 &  \sqrt[]{2} & 0 \\
               i & 0 &  -i \\
             \end{array}
             \right),
  \end{eqnarray}
and by defining $H_{0}^{\prime}(t)=V^{\dag}H_1^{\prime}(t)V$, we can transform $H_1^{\prime}(t)$ into
\begin{eqnarray}\label{eqb15}
H_{0}^{\prime}(t)=\Omega_x(t)J_{x}+\Omega_y(t)J_{y}+0J_z,
\end{eqnarray}
 where $J_{x}$, $J_{y}$, $J_{z}$ are angular-momentum operators for spin 1. It's obvious that $H_{0}^{\prime}(t)$ possesses the same form as $H_0(t)$ in Eq.~(1) as long as we replace $\sigma_x,\sigma_y,\sigma_z$ with $J_{x}$, $J_{y}$, $J_{z}$, respectively. Thus, if we start from $H_{0}^{\prime}(t)$, a similar procedure to inverse designing Hamiltonian $H_{0}^{\prime}(t)$ can be carried out in a three-level system.

Imitating the constructing procedure in Eqs.~(2)-(5), in the three-level system the transformation operator is set as
\begin{eqnarray}\label{eqb16}
    R^{\prime}(\alpha,\theta,\beta)
                    =e^{\frac{i\alpha}{2}J_{z}}e^{-i\theta J_{y}}e^{\frac{i\beta}{2} J_{z}}.
  \end{eqnarray}
Then similar to Eqs.~(4) and (5), in the picture defined by $R^{\prime\dag}$, we will get $H^{\prime}(t)=R^{\prime\dag}H^{\prime}_0(t)R^{\prime}+i\partial_t{R^{\prime\dag}}R^{\prime}$ and $\textrm{U}_O^{\prime}(t)=R^{\prime}(t)\textrm{U}_R^{\prime}(t)R^{\prime\dag}(0)$, here $\textrm{U}_O^{\prime}(t)$ and $\textrm{U}_R^{\prime}(t)$ are the evolution operators in pictures $V$ and $R^{\prime\dag}$, respectively. By careful calculation, we can deduce
 \begin{eqnarray}\label{eqb3}
  H^{\prime}(t)            &=&J_{x}[\Omega_{x}\cos\frac{\alpha}{2}\cos\frac{\beta}{2}\cos\theta-\Omega_{x}\sin\frac{\alpha}{2}\sin\frac{\beta}{2}
               \cr
               &-&\Omega_{y}\sin\frac{\alpha}{2}\cos\frac{\beta}{2}\cos\theta-\Omega_{y}\cos\frac{\alpha}{2}\sin\frac{\beta}{2}
               \cr
               &+&\dot{\theta}\sin\frac{\beta}{2}-\frac{\dot{\alpha}}{2}\cos\frac{\beta}{2}\sin\theta]
                 \cr
               &+&J_{y}[\Omega_{x}\cos\frac{\alpha}{2}\sin\frac{\beta}{2}\cos\theta+\Omega_{x}\sin\frac{\alpha}{2}\cos\frac{\beta}{2}
               \cr
               &-&\Omega_{y}\sin\frac{\alpha}{2}\sin\frac{\beta}{2}\cos\theta+\Omega_{y}\cos\frac{\alpha}{2}\cos\frac{\beta}{2}
               \cr
               &-&\dot{\theta}\cos\frac{\beta}{2}-\frac{\dot{\alpha}}{2}\sin\frac{\beta}{2}\sin\theta]
               \cr
               &+&J_{z}[\Omega_{x}\cos\frac{\alpha}{2}\sin\theta-\Omega_{y}\sin\frac{\alpha}{2}\sin\theta
               \cr
               &+&\frac{\dot{\beta}}{2}+\frac{\dot{\alpha}}{2}\cos\theta].
\end{eqnarray}
Similarly, as described in the two-level system, we can choose apposite $R^{\prime}(t)$ and $H^{\prime}(t)$ to inverse design Hamiltonian $H^{\prime}_0(t)$, then, Hamiltonian $H_1^{\prime}(t)=VH^{\prime}_0(t)V^{\dag}$ that evolves the system is determined accordingly.
That is, the inverse designation of a Hamiltonian $H_1^{\prime}(t)$ is completed.
 Note that, here in the Hamiltonian $H^{\prime}_0(t)$, the term with $J_z$ is undesired, because the inverse transformation $VJ_zV^{\dag}$ will lead to an additional transition between the levels $|1\rangle$ and $|3\rangle$, which is usually electric dipole forbidden in some systems \cite{ChenXPRL10105}.

\section{ FAST STATE ENGINEERING IN NV CENTERS VIA DESIGNED SHORTCUTS }\label{section:III}

\begin{figure}
\scalebox{0.2}{\includegraphics{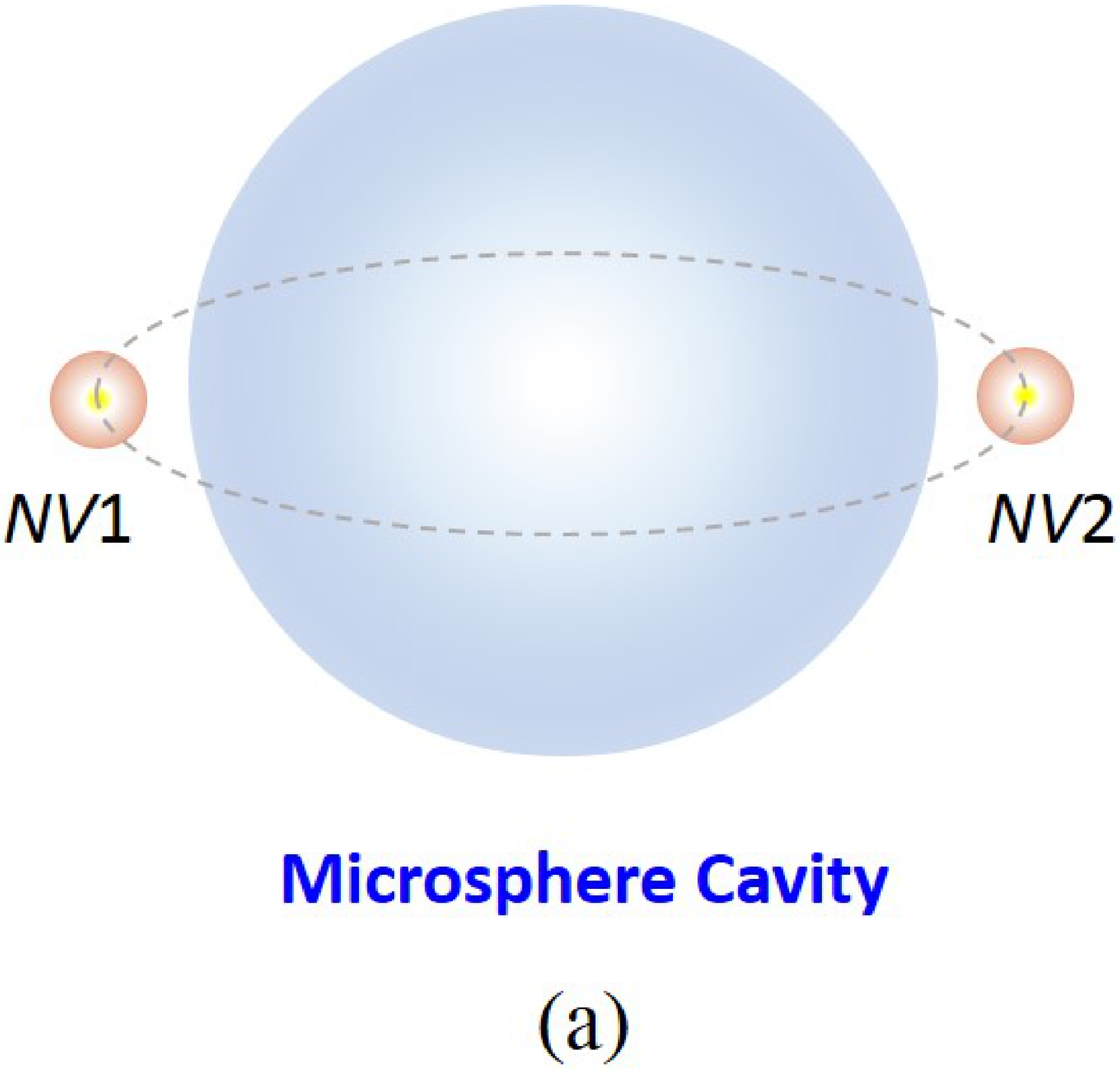}}
\scalebox{0.2}{\includegraphics{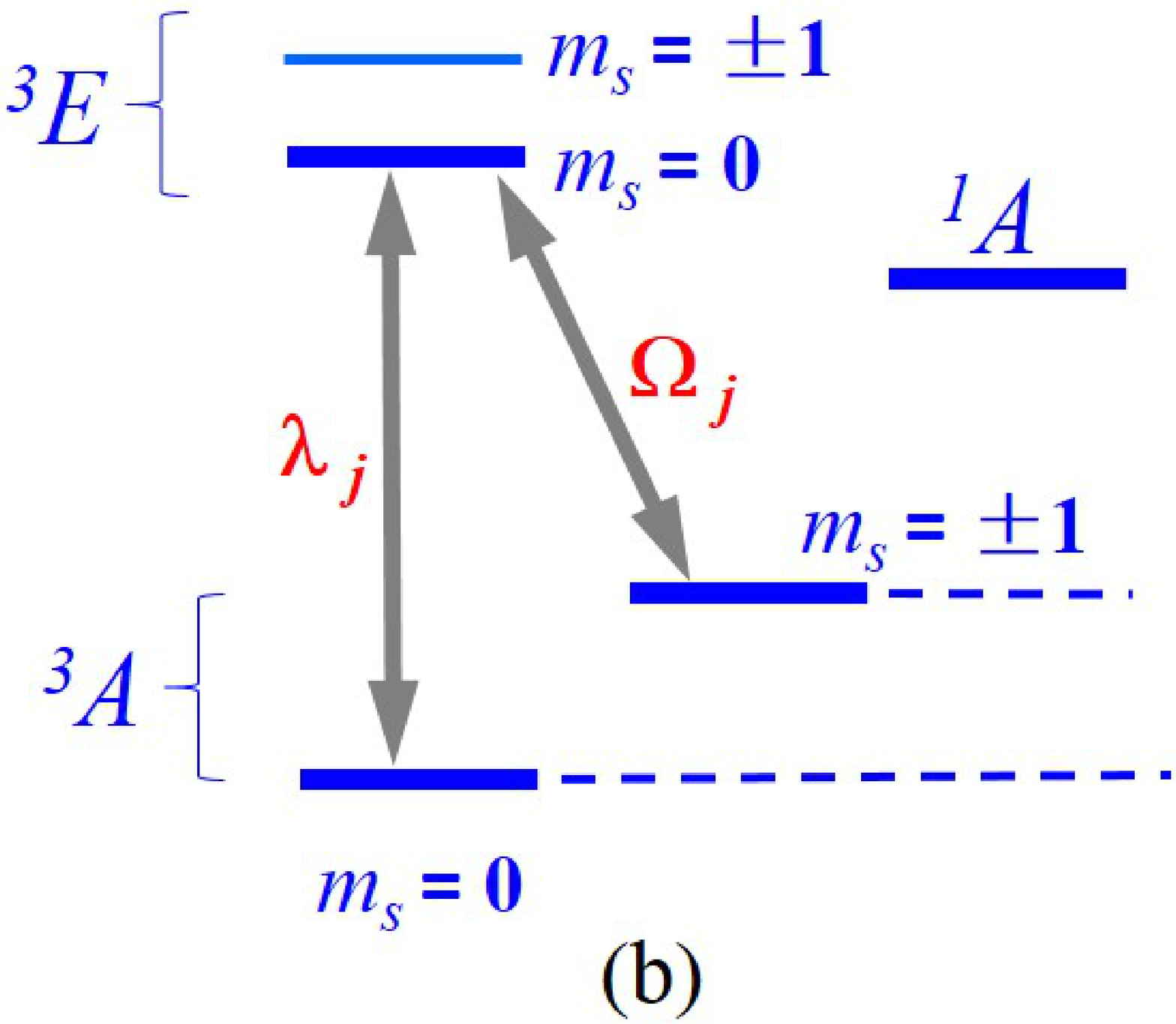}}
\caption
   {\label{fig3}
(a) Schematic setup of the fused-silica microsphere cavity, where two identical NV centers in diamond are attached around the equator of the cavity.
(b) Level diagram for an NV center, with $\lambda$~($\Omega$) the coupling strength between the NV center and the WGM~(laser pulse). Quantum information is encoded in the spin states $m_s=0$ and $m_s=-1$.
}
\end{figure}

In this section, we will take advantages of the constructed shortcuts to present an experimentally feasible scheme for the implementation of population transfer between distant NV centers in diamond coupled to the WGM of a microsphere cavity \cite{BarclayAPL0995,DujfNJP1012,ParkNL0606}. In the cavity, the lowest-order WGM, corresponding to the light traveling around the equator of the microsphere \cite{Buck0367}, offers exceptional mode properties for reaching strong light-matter coupling \cite{BarclayAPL0995,DujfNJP1012,ParkNL0606}. The setup we consider and the energy level configuration of the NV centers are sketched in Fig.~3. Two negatively charged NV centers are separately positioned near the equator of a high-$Q$ microsphere cavity, in which the distance is much larger than the wavelength of the WGM and the direct coupling between NV centers is negligible. The NV centers can be modeled as $\Lambda$-type three level systems \cite{ToganNature10466,WoltersPRA1489,DujfAPL1096}, where the states $|^{3}A,m_s=0\rangle$ and $|^{3}A,m_s=-1\rangle$ serve as qubit states $|g\rangle$ and $|f\rangle$, respectively, the state $|^{3}E,m_s=0\rangle$ is labeled by $|e\rangle$ and the metastable state $^{1}A$ is left aside for not fully understood \cite{MansonPRB0674}. In our case, the transition $|g\rangle\leftrightarrow|e\rangle$ is resonantly coupled to the WGM with coupling constant $\lambda$ and $|f\rangle\leftrightarrow|e\rangle$ is resonantly driven through a time-dependent laser pulse with Rabi frequency $\Omega(t)$ \cite{DujfNJP1012,SantoriPRL0697}. Using the rotating-wave approximation~(RWA), in the interaction picture, the interacting Hamiltonian is written as $(\hbar=1)$
\begin{eqnarray}\label{eqb17}
     H_I&=&H_L+H_c,
     \cr
     H_L&=&\sum_{j=1,2}\Omega_j(t)|e\rangle_j\langle f|+H.c.,~~
     \cr
     H_c&=&\sum_{j=1,2}\lambda_j|e\rangle_j\langle g|\hat{a}+H.c.,
  \end{eqnarray}
where subscript $j$ denotes the $j$th NV center, $\hat{a}~(\hat{a}^\dag)$ is the annihilation~(creation) operator of the WGM field. For simplicity, we adopt $\lambda_j=\lambda$ in the following. Moreover,
we assume the initial state of the system is
$|\psi_{0}\rangle=|f\rangle_1|g\rangle_2|0\rangle_c$. Defining the excited number operator of the system as $N_e=\sum_j(|e\rangle_j\langle e|+|f\rangle_j\langle f|)+a^{\dag}a$, one can obtain $[N_e,H_I]=0$ and $\langle\psi_{0}|N_e|\psi_0\rangle=1$. Therefore, the system will
evolve within a single-excitation subspace spanned by
\begin{eqnarray}\label{eqc18}
  |\psi_{1}\rangle&=&|f\rangle_1|g\rangle_2|0\rangle_c, ~~~~ |\psi_{2}\rangle=|e\rangle_1|g\rangle_2|0\rangle_c,
  \cr
  |\psi_{3}\rangle&=&|g\rangle_1|g\rangle_2|1\rangle_c, ~~~~
  |\psi_{4}\rangle=|g\rangle_1|e\rangle_2|0\rangle_c,
  \cr
  |\psi_{5}\rangle&=&|g\rangle_1|f\rangle_2|0\rangle_c.
    \end{eqnarray}
Moreover, the eigenstates of $H_c$ can be described as
\begin{eqnarray}\label{eqc19}
  |\phi_{1}\rangle&=&\frac{1}{\sqrt{2}}(-|\psi_{2}\rangle+|\psi_{4}\rangle),
  \cr
  |\phi_{2}\rangle&=&\frac{1}{2}(|\psi_{2}\rangle+\sqrt{2}|\psi_{3}\rangle+|\psi_{4}\rangle),  \cr
  |\phi_{3}\rangle&=&\frac{1}{2}(|\psi_{2}\rangle-\sqrt{2}|\psi_{3}\rangle+|\psi_{4}\rangle), \end{eqnarray}
with corresponding eigenvalues $\varepsilon_1=0,\varepsilon_2=\sqrt{2}\lambda$, and $\varepsilon_3=-\sqrt{2}\lambda$, respectively.
Rewriting the Hamiltonian in Eq.~(22) with the eigenvectors of $H_c$ we obtain
 \begin{eqnarray}\label{eqb17}
     H_I&=&H_{L,r}+H_{c,r},
     \cr
     H_{c,r}&=&\sum_{m=1}^{3}\varepsilon_m|\phi_m\rangle\langle\phi_m|,
     \cr
     H_{L,r}&=&-\frac{\Omega_1(t)}{\sqrt{2}}|\phi_1\rangle\langle\psi_1|+\frac{\Omega_1(t)}{2}|\phi_2\rangle\langle\psi_1|
     \cr
     &+&\frac{\Omega_1(t)}{2}|\phi_3\rangle\langle\psi_1|+\frac{\Omega_2(t)}{\sqrt{2}}|\phi_1\rangle\langle\psi_5|
     \cr
     &+&\frac{\Omega_2(t)}{2}|\phi_2\rangle\langle\psi_5|+\frac{\Omega_2(t)}{2}|\phi_3\rangle\langle\psi_5|+H.c..~~
  \end{eqnarray}
Performing a unitary transformation $U=e^{iH_{c,r}t}$ on $H_I$, we obtain
\begin{eqnarray}\label{eqb17}
     H_{I,e}&=&UH_IU^{\dag}+i\dot{U}U^{\dag}
     \cr
     &=&-\frac{\Omega_1(t)}{\sqrt{2}}|\phi_1\rangle\langle\psi_1|+\frac{\Omega_1(t)}{2}e^{i\sqrt{2}\lambda t}|\phi_2\rangle\langle\psi_1|
     \cr
     &+&\frac{\Omega_1(t)}{2}e^{-i\sqrt{2}\lambda t}|\phi_3\rangle\langle\psi_1|+\frac{\Omega_2(t)}{\sqrt{2}}|\phi_1\rangle\langle\psi_5|
     \cr
     &+&\frac{\Omega_2(t)}{2}e^{i\sqrt{2}\lambda t}|\phi_2\rangle\langle\psi_5|+\frac{\Omega_2(t)}{2}e^{-i\sqrt{2}\lambda t}|\phi_3\rangle\langle\psi_5|
     \cr
     &+&H.c..~~
  \end{eqnarray}
 On the condition $\sqrt{2}\lambda\gg\Omega_j(t)$, the terms in $H_{I,e}$ with high oscillating frequency $\sqrt{2}\lambda$ can be ignored. So the effective Hamiltonian governing the evolution can be given as
\begin{eqnarray}\label{eqb20}
     H_{eff}=-\frac{\Omega_1(t)}{\sqrt{2}}|\phi_1\rangle\langle\psi_1|+\frac{\Omega_2(t)}{\sqrt{2}}|\phi_1\rangle\langle\psi_5|+H.c..
  \end{eqnarray}
If we regard $|\psi_1\rangle$ and $|\psi_5\rangle$ as two ground states and $|\phi_1\rangle$ as an excited state, replacing $|\psi_1\rangle$ with $|\psi_0\rangle$,
it's obvious that $H_{eff}$ can be read as
 \begin{eqnarray}\label{eqb14}
     H_{eff}= \frac{1}{\sqrt{2}}
            \left(
             \begin{array}{cccc}
               0 & -\Omega_1(t) &  0 \\
               -\Omega_1(t) &  0 & \Omega_2(t) \\
               0 & \Omega_2(t) &  0 \\
             \end{array}
            \right).
  \end{eqnarray}
   Replacing $-\frac{\Omega_1(t)}{\sqrt{2}},\frac{\Omega_2(t)}{\sqrt{2}}$ with $\Omega_x,\Omega_y$ respectively, it is obvious that
 $H_{eff}$ possesses the same form as the Hamiltonian in
Eq.~(17). Thus we could apply the procedure discussed in Sec.~II to achieve quantum state transfer in the two NV centers.

Here, $H^{\prime}(t)$ is defined as $H^{\prime}(t)=R^{\prime\dag}H_{eff} R^{\prime}+i\partial_tR^{\prime\dag}R^{\prime}$. We choose the case of $\beta=0$ and $H^{\prime}(t)=f^{\prime}_z(t)J_z$ to construct the shortcut. According to Eq.~(21), setting the coefficients of $J_x, J_y$ to zero, it is easy to obtain
\begin{eqnarray}\label{eqb21}
    -\frac{\Omega_1(t)}{\sqrt{2}}\equiv\widetilde{\Omega}_1&=&\dot{\theta}\sin\frac{\alpha}{2}+\frac{\dot{\alpha}}{2}\cos\frac{\alpha}{2}\tan\theta,~~~
    \cr
    \frac{\Omega_2(t)}{\sqrt{2}}\equiv\widetilde{\Omega}_2&=&\dot{\theta}\cos\frac{\alpha}{2}-\frac{\dot{\alpha}}{2}\sin\frac{\alpha}{2}\tan\theta,
  \end{eqnarray}
and
\begin{eqnarray}\label{eqb22}
     f^{\prime}_z(t)=\frac{\dot{\alpha}}{2\cos\theta}.
  \end{eqnarray}
 When $\beta=0$, according to Eq.~(20), the unitary transformation is $R^{\prime}=e^{\frac{i\alpha}{2}J_z}e^{-i\theta J_y}$. Since $H^{\prime}(t)=f^{\prime}_z(t)J_z$,
 the evolution operator in the picture $R^{\prime}$ is $\textrm{U}_R^{\prime}(t)=e^{-\frac{i\delta^{\prime}_z}{2}J_z}$ with $\delta^{\prime}_z=\int^{t}_0\frac{\dot{\alpha}}{\cos\theta}dt^{\prime}$.
  Therefore,
 the evolution operator in picture $V$ will be
\begin{eqnarray}\label{eqb23}
    \textrm{U}_O^{\prime}(t)=R^{\prime}(t)\textrm{U}_R^{\prime}(t)=e^{\frac{i\alpha}{2}J_z}e^{-i\theta J_y}e^{-\frac{i\delta^{\prime}_z}{2}J_z}.
  \end{eqnarray}

For the population transfer process, we take $|\psi_0\rangle=(1,0,0)^T$ as the initial state and $|\psi_5\rangle=(0,0,1)^T$ as the target state. In the picture defined by $V$, it means we should complete the process $V^{\dag}|\psi_0\rangle=\frac{1}{\sqrt{2}}(1,0,1)^T$ to $V^{\dag}|\psi_5\rangle=\frac{1}{\sqrt{2}}(1,0,-1)^T$ (up to a global phase).
Then through detailed calculation on $\textrm{U}_V(t)$, we can deduce
  $\frac{1}{\sqrt{2}}(-\sin\theta e^{\frac{i\alpha}{2}}, \cos\theta, \sin\theta e^{-\frac{i\alpha}{2}})^T$ as the moving state of the system. To obtain the population inversion, the boundary conditions can be set as
  \begin{eqnarray}\label{eqb24}
    \alpha(0)=\pi,~~~ \alpha(T)=0,~~~  \theta(0)=\frac{\pi}{2},~~~  \theta(T)=\frac{\pi}{2}.
  \end{eqnarray}
To make the external driving fields could be smoothly turned on and turned off, we would like to add the boundary conditions
 \begin{eqnarray}\label{eqb25}
    \dot{\theta}(0)=\dot{\theta}(T)=0,~~~ \dot{\mu}(0)=\dot{\mu}(T)=0,
  \end{eqnarray}
 where $\dot{\mu}$ is defined as $\dot{\mu}=-\frac{\dot{\alpha}}{2\cos\theta}$. Considering Eq.~(32) and Eq.~(33), parameters $\theta$ and $\dot{\mu}$ can be designed as
 \begin{eqnarray}\label{eqb26}
\theta&=&\frac{\pi}{2}-\frac{A}{2}(\frac{t}{T})^2[1-2(\frac{t}{T})+(\frac{t}{T})^2],
\cr
\dot{\theta}&=&A\frac{t}{T^2}(1-\frac{t}{T})(2\frac{t}{T}-1),
\cr
\dot{\mu}&=&B\frac{t}{T}(1-\frac{t}{T}),
  \end{eqnarray}
where $A$ and $B$ are time-independent adjustable parameters. We set
$A\in(0,32)$ to avoid singularity of the expression and optimize the
amplitude for each pulse. Note that $A,B$ should be chosen such that
\begin{eqnarray}\label{eqb27}
    \alpha (0)=\int_{0}^{T}2\dot{\mu}(t)\cos\theta(t) dt=\pi.
  \end{eqnarray}

 According to Eqs.~(29) and (34), the amplitudes of pulses $\widetilde{\Omega}_{1},\widetilde{\Omega}_{2}$ and the maximal population $P_{Imax}$ of the intermediate state $|\phi_1\rangle$ can be controlled by $A$ and $B$.
Some samples of $A$ and $B$ with corresponding $P_{Imax}$, $\widetilde{\Omega}_{max}T$ are given in Table I.
\begin{center}
{\bf Table I. Samples of $A,B$ with corresponding $P_{Imax}$, $\widetilde{\Omega}_{max}T$.}\\
{\small
  \begin{tabular}{cccc}
  \hline
  \hline
$ A $ \ \ \ \ \ \ \ \ \ \ \ \ \ \  & $B$ \ \ \ \ \ \ \ \ \ \ \ \ \ \  &$P_{Imax}$\ \ \ \ \ \ \ \ \ \ \ \ \ \  &$\widetilde{\Omega}_{max}T$ \\
\hline
$9.5\ \ \ \ \ \ \ \ \ \ \ \ \ \ $ & $46.7738\ \ \ \ \ \ \ \ \ \ \ \ \ \ $ & $0.085\ \ \ \ \ \ \ \ \ \ \ \ \ \ $ & $9.629$\\
$10\ \ \ \ \ \ \ \ \ \ \ \ \ \ $ & $44.4844\ \ \ \ \ \ \ \ \ \ \ \ \ \ $ &$0.094\ \ \ \ \ \ \ \ \ \ \ \ \ \ $ & $9.103$\\
$10.5\ \ \ \ \ \ \ \ \ \ \ \ \ \ $ & $42.4155\ \ \ \ \ \ \ \ \ \ \ \ \ \ $ &$0.104\ \ \ \ \ \ \ \ \ \ \ \ \ \ $ & $8.625$\\
$11\ \ \ \ \ \ \ \ \ \ \ \ \ \ $ & $40.5370\ \ \ \ \ \ \ \ \ \ \ \ \ \ $ & $0.113\ \ \ \ \ \ \ \ \ \ \ \ \ \ $ & $8.188$\\
$11.5\ \ \ \ \ \ \ \ \ \ \ \ \ \ $ & $38.8242\ \ \ \ \ \ \ \ \ \ \ \ \ \ $ &$0.123\ \ \ \ \ \ \ \ \ \ \ \ \ \ $ & $7.788$\\
$12\ \ \ \ \ \ \ \ \ \ \ \ \ \ $ & $37.2564\ \ \ \ \ \ \ \ \ \ \ \ \ \ $ &$0.134\ \ \ \ \ \ \ \ \ \ \ \ \ \ $ & $7.420$\\
\hline
 \hline
\end{tabular}
}
\end{center}

As seen from Table I, when $A$ decreases, the population of intermediate state $|\phi_1\rangle$ decreases, while the value of $\widetilde{\Omega}_{max}T$ increases, that means if the pulses' amplitude $\widetilde{\Omega}_{max}$ is fixed, the interaction time $T$ will increase. Since
 the intermediate states are lossy states, large populations are undesired. At the same time, in view of low energy consuming, we do not expect a large amplitude of pulse. Thus, it is better to choose a suitable $A$ so that both the population of intermediate state and the interaction time can be restricted in a desired range. Here we choose $A=11, B=40.537$ to fulfill the task. With these chosen parameters, $\widetilde{\Omega}_1$ and $\widetilde{\Omega}_2$ are plotted by solid-blue lines in Figs.~4(a) and 4(b), respectively. For the convenience of experimental realization, we engineer Gaussian-shaped pulses to fit $\widetilde{\Omega}_1$ and $\widetilde{\Omega}_2$, and obtain two substituted pulses $\overline{\Omega}_1$ and $\overline{\Omega}_2$ plotted by dotted-red lines in Figs.~4(a) and 4(b), respectively. The expressions of $\overline{\Omega}_1$ and $\overline{\Omega}_2$ are
\begin{eqnarray}\label{eqb28}
    \overline{\Omega}_q&=&\zeta_{q}e^{-(\frac{t-\tau_{q}}{\sigma_{q}})^2} ~~(q=1,2),
    \cr
    \zeta_{1}&=&-8.283/T,~~\tau_{1}=0.6277T,~~\sigma_{1}=0.299T,
    \cr
   \zeta_{2}&=&8.283/T,~~\tau_{2}=0.3722T,~~\sigma_{2}=0.299T.
  \end{eqnarray}

Next, we will investigate the performance of the protocol via numerical simulation. We define the fidelity of the target state $|\psi_5\rangle$ as $F(t)=|\langle\psi_5|\rho(t)|\psi_5\rangle|$, where $\rho(t)$ is the density operator of the system. Note that, to execute the process successfully, we need to ensure that
 $\sqrt{2}\lambda\gg \Omega_j(t)$ as it is the condition to obtain the effective Hamiltonian $H_{eff}$, so before further discussions with the original Hamiltonian $H_I$, it is better to choose a suitable value for the coupling constant $\lambda$.
In the present protocol, according to Eq.~(29), the condition $\sqrt{2}\lambda\gg \Omega_j(t)$ can be replaced by $\lambda\gg \overline{\Omega}_q(t)$ and the pulse amplitude is $\overline{\Omega}_0=\max{(\overline{\Omega}_1(t),\overline{\Omega}_2(t))}\approx8.3/T$. We plot the final fidelity $F(T)$ versus $\lambda$ in Fig.~5(a).
As shown in Fig.~5(a), the final fidelity $F(T)$ is almost 1 when $\lambda \geq 20/T$, which means that even if $\lambda\gg \overline{\Omega}_q(t)$ is not satisfied, the target state $|\psi_5\rangle$ can also be obtained. However, if $\lambda$ is not large enough to satisfy $\lambda\gg \overline{\Omega}_q(t)$, the system may evolve along an unknown path not governed by the effective Hamiltonian. As a result, the population of each intermediate state cannot be forecasted as before, thus the system may suffer more from dissipations and finally have a relatively poor performance when decoherence is taken into considered. On the other hand, for a relative higher evolution speed, the value of $\lambda T$ should not be too large, as $\lambda$ generally has an upper limit in real experiments.
Therefore, to make the protocol with both high speed and robustness against dissipations, we adopt $\lambda=30/T~(\overline{\Omega}_0/\lambda\approx0.28)$ for demonstration.
 Based on Eq.~(36), we plot the populations versus $t/T$ in Fig.~5(b). As shown in Fig.~5(b), the population of $|\psi_5\rangle$ ($P_5$) increases from 0 to 1 during the evolution, which shows the
present protocol could complete the population transfer successfully. Moreover, the populations of the intermediate states keep lower than 0.12.
\begin{figure}
  \scalebox{0.35}{\includegraphics{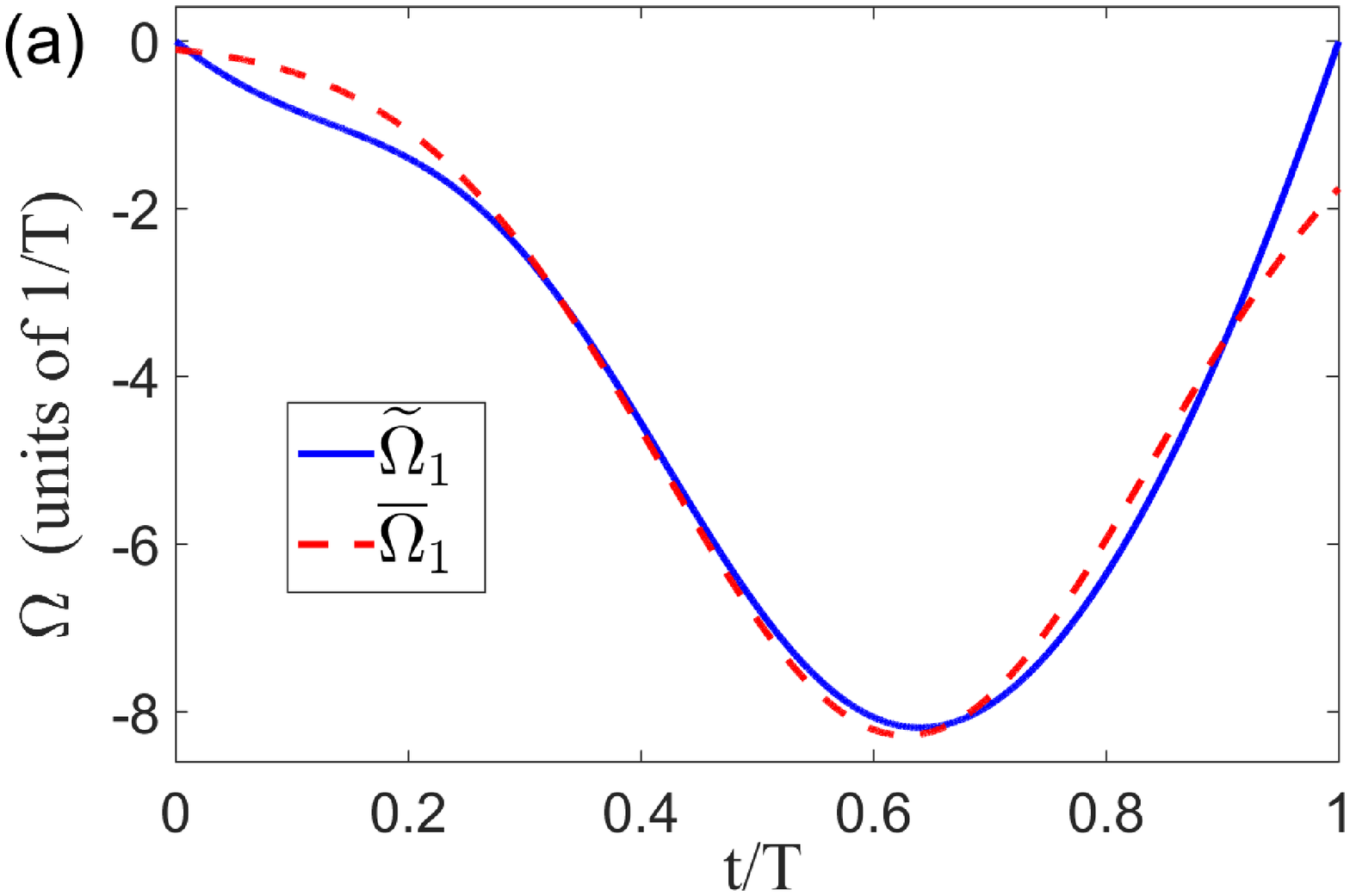}}
  \scalebox{0.35}{\includegraphics{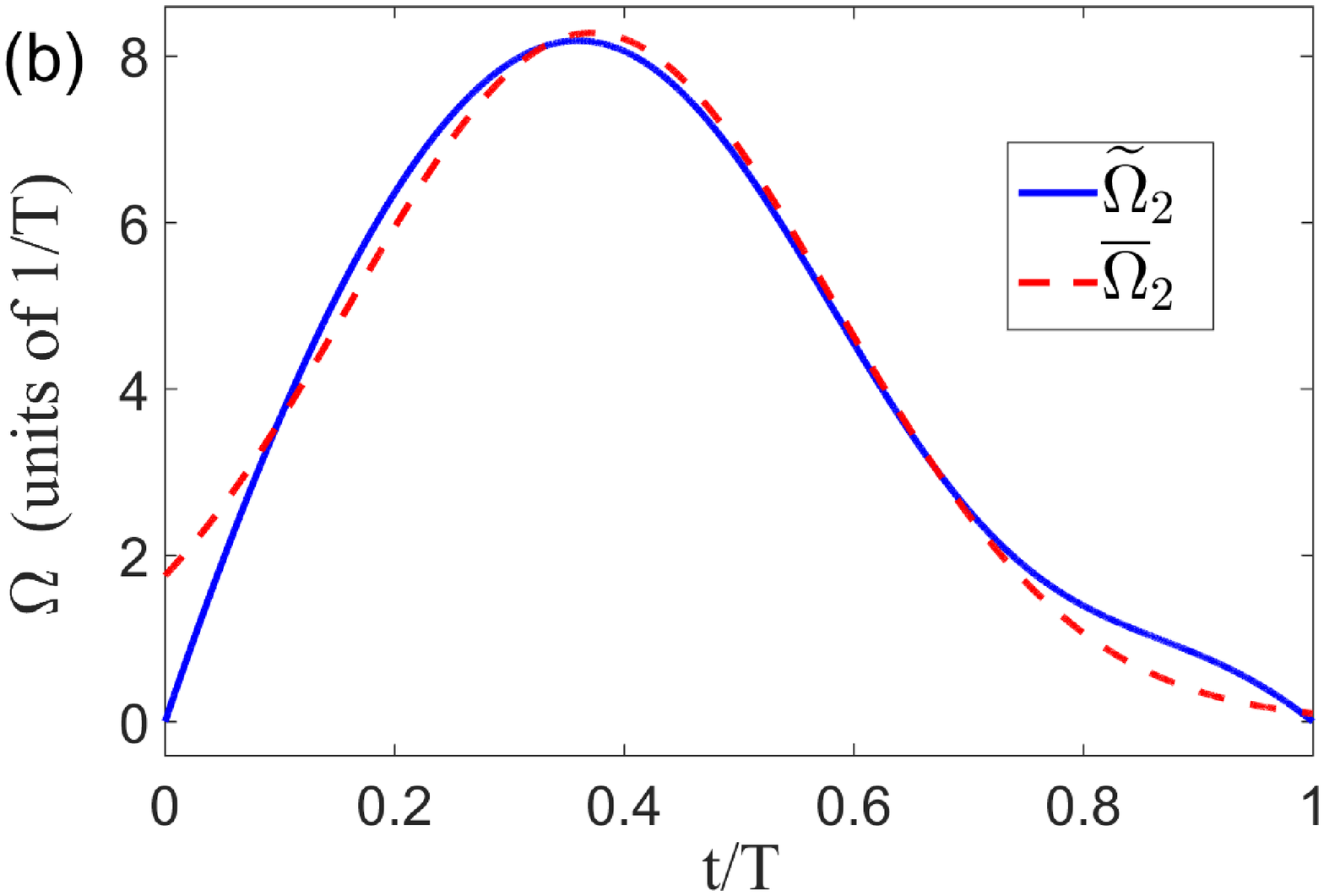}}
  \caption{
         (a) $\widetilde{\Omega}_1(t)$ and $\overline{\Omega}_1(t)$ versus $t/T$.
         (b) $\widetilde{\Omega}_2(t)$ and $\overline{\Omega}_2(t)$ versus $t/T$.
         }
 \label{fig4}
\end{figure}
\begin{figure}
 \scalebox{0.35}{\includegraphics{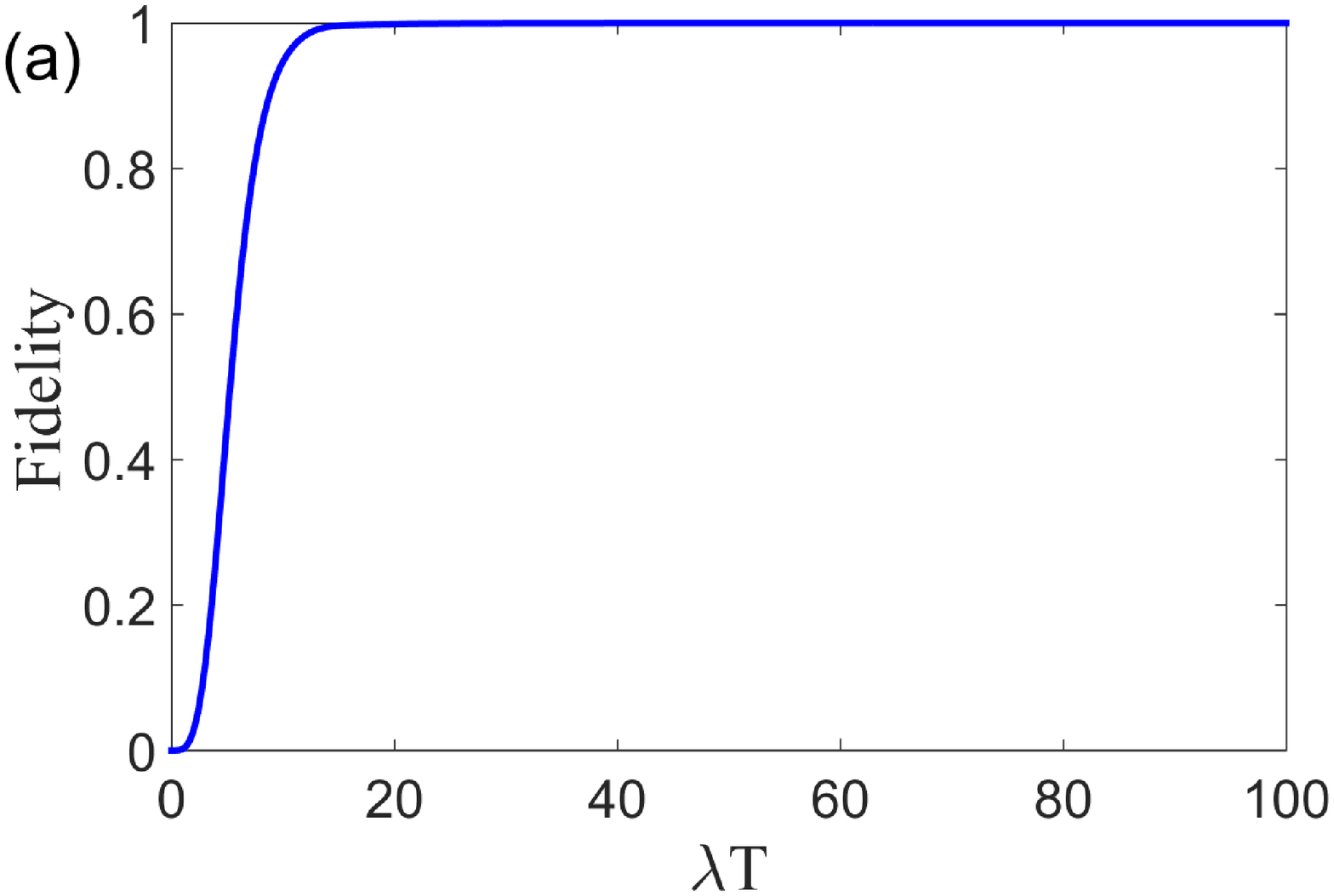}}
\scalebox{0.35}{\includegraphics{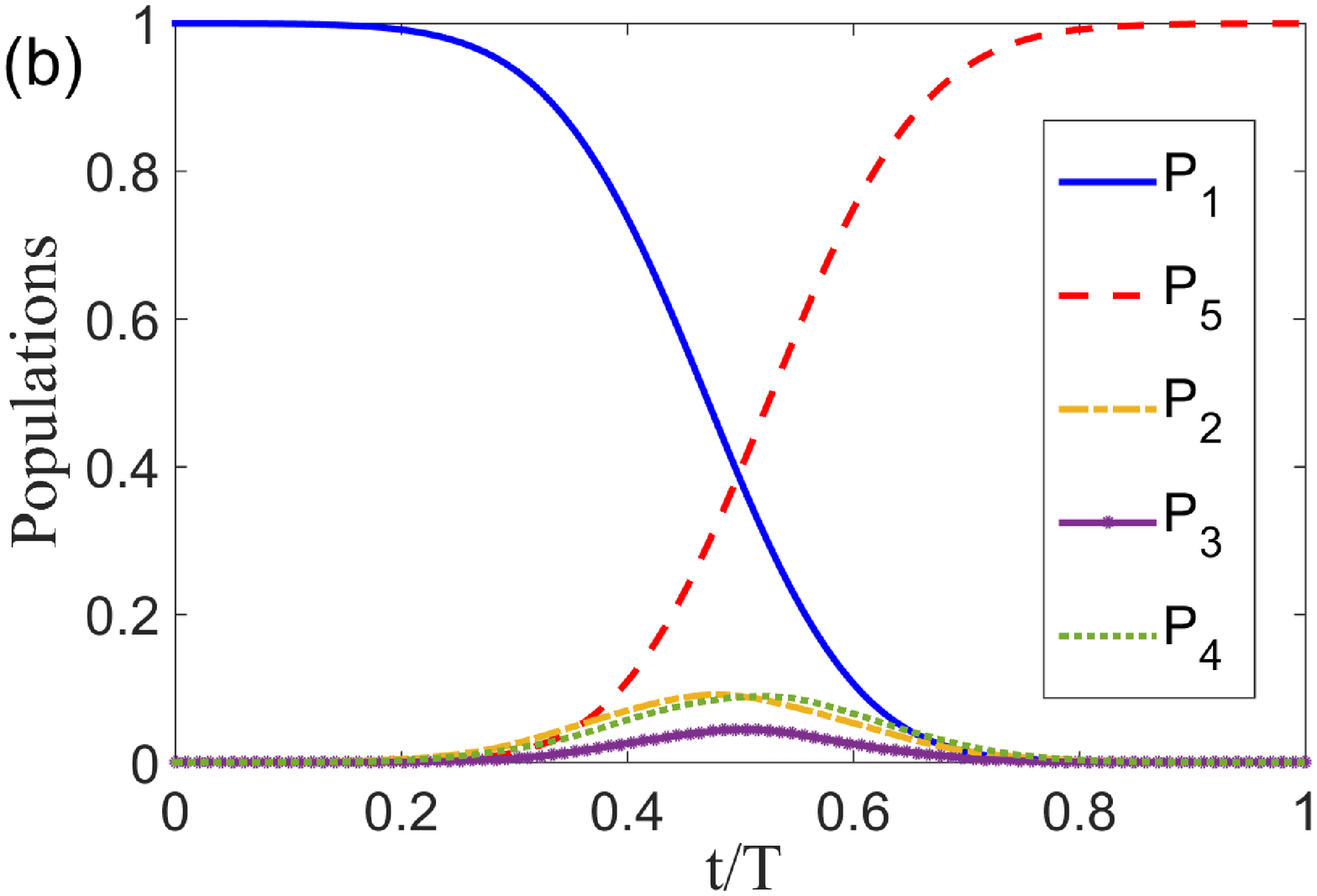}}
 \caption{
         (a) The final fidelity $F(T)$ of state $|\psi_5\rangle$ versus $\lambda T$.
         (b) Populations of $|\psi_1\rangle$~($P_1$, the solid blue line), $|\psi_2\rangle$~($P_2$, the dashed-dotted yellow line), $|\psi_3\rangle$~($P_3$, the solid purple line), $|\psi_4\rangle$~($P_4$, the dotted green line), $|\psi_5\rangle$~($P_5$, the dashed red line) versus $t/T$.
         }
 \label{fig5}
\end{figure}
\begin{figure}
 \scalebox{0.35}{\includegraphics {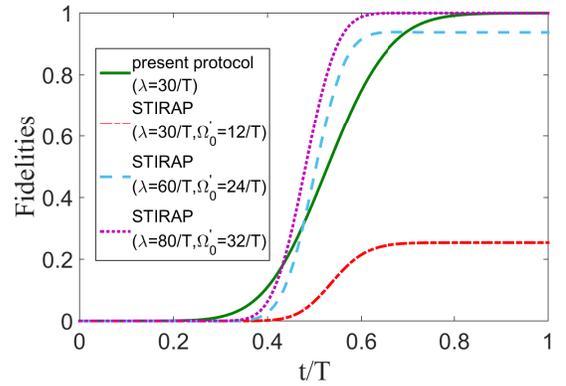}}
 \caption{
         The fidelities of the target state $|\psi_5\rangle$ versus $t/T$ with different methods: solid green line, using the present method with $\lambda=30/T$; dashdotted
red line, STIRAP with $\lambda = 30/T$ and $\Omega^{\prime}_0 = 12/T$ ; dashed
blue line, STIRAP with $\lambda = 60/T$ and $\Omega^{\prime}_0 = 24/T$; and dotted purple
line, STIRAP with $\lambda = 80/T$ and $\Omega^{\prime}_0 = 32/T$.
         }
 \label{fig6}
\end{figure}

To show the fastness of the current protocol over an adiabatic protocol, we apply the stimulated Raman adiabatic passage (STIRAP) to achieve the quantum state transfer. Based on STIRAP, the time-dependent laser pulses are usually designed as Gaussian functions in the forms of
\begin{eqnarray}\label{eqb29}
    \Omega^{\prime}_1=\Omega^{\prime}_0e^{-(\frac{t-t_1}{t_c})^2},~~~
    \Omega^{\prime}_2=\Omega^{\prime}_0e^{-(\frac{t-t_2}{t_c})^2},
     \end{eqnarray}
where $t_1, t_2,$ and $t_c$ are three related parameters and can be
selected as $t_1=0.54T,~t_2=0.4T,~t_c=0.14T$ \cite{DujfNJP1012}. Based
on Eq.~(37), we plot Fig.~6 to compare the speed of the present
protocol with that of STIRAP. As shown in Fig.~6, the fidelity of
the present protocol can reach 1 at $t=T$ (solid green line) while
with the same condition for STIRAP ($\Omega^{\prime}_0=12/T,~
\lambda=30/T$; dash-dotted red line), the fidelity is only about
0.254 due to bad violation of the adiabatic condition. If the pulse
amplitude $\Omega^{\prime}_0$ is increased to $32/T$ and $\lambda$
is $80/T$ (see dotted purple line of Fig.~6), for the STIRAP method,
the fidelity can approach 1. However, in this case, the laser
amplitude $\Omega^{\prime}_0=32/T$ is much larger than the one
($\overline{\Omega}_0=8.3/T$) of the present protocol. As is known,
if a relative high evolution speed is desired, the product of the
pulse amplitude and the total interaction time $T$ is the smaller
the better, because when the laser amplitude has a fixed
value, the one with smaller product will have less evolution time.
Therefore, the speed of the present protocol to obtain the target
state is much faster when compared with that of the STIRAP method.
\begin{figure}
 \scalebox{0.35}{\includegraphics {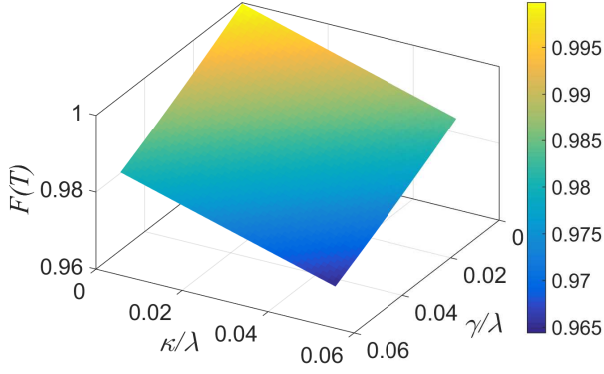}}
 \caption{
         The final fidelity $F(T)$ versus $\kappa/\lambda$ and $\gamma/\lambda$.
         }
 \label{fig7}
\end{figure}
\begin{figure}
\scalebox{0.4}{\includegraphics{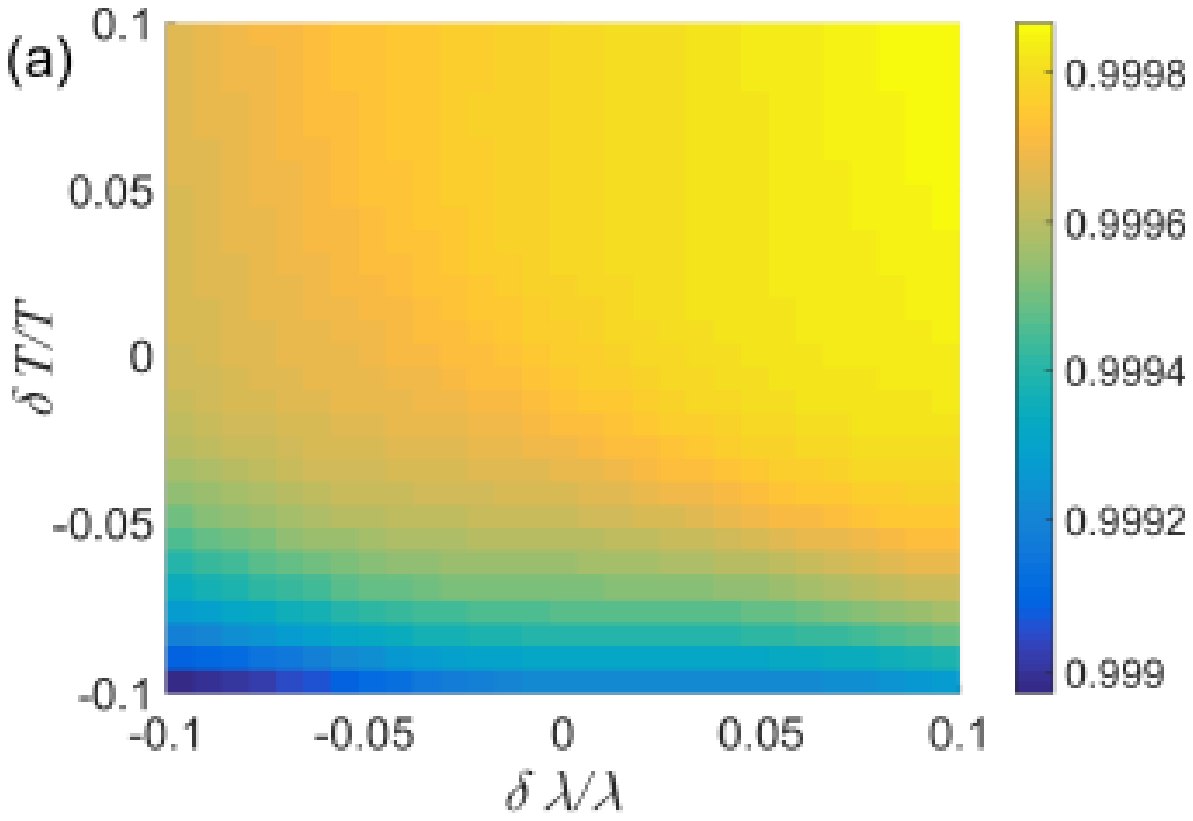}}
\scalebox{0.4}{\includegraphics{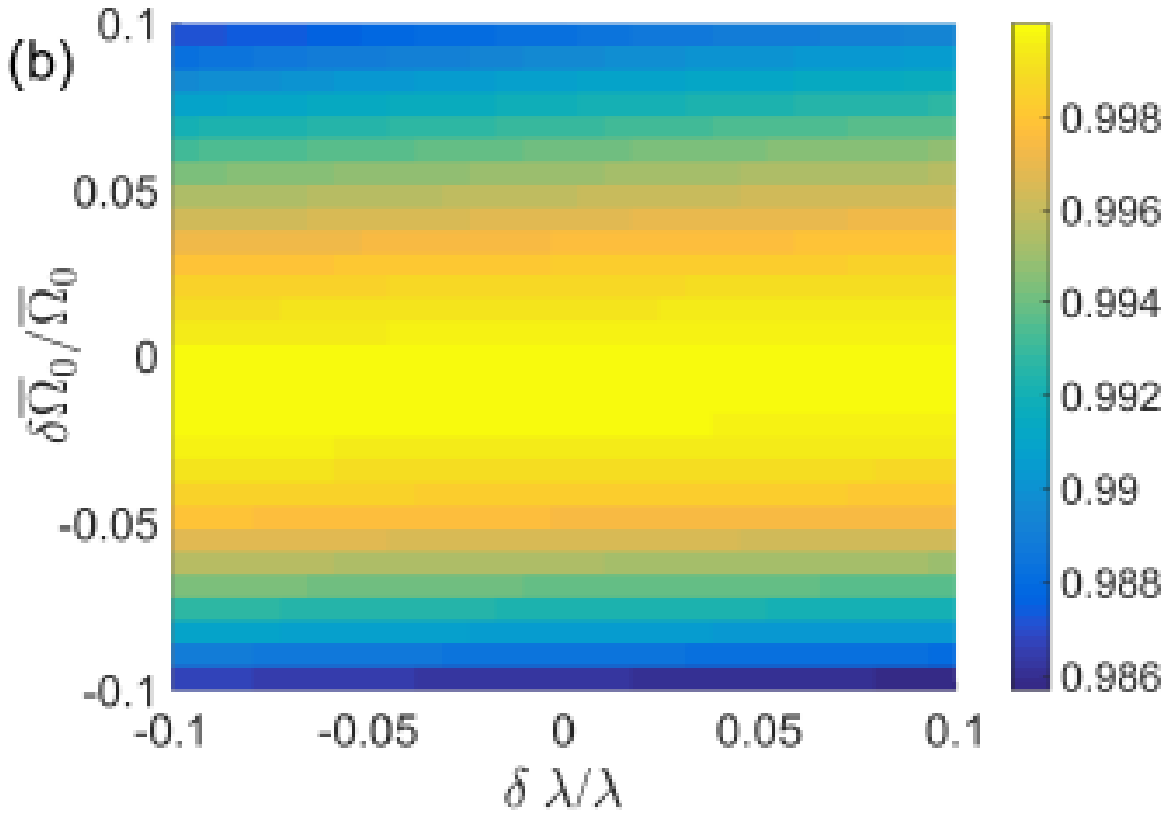}}
\scalebox{0.4}{\includegraphics{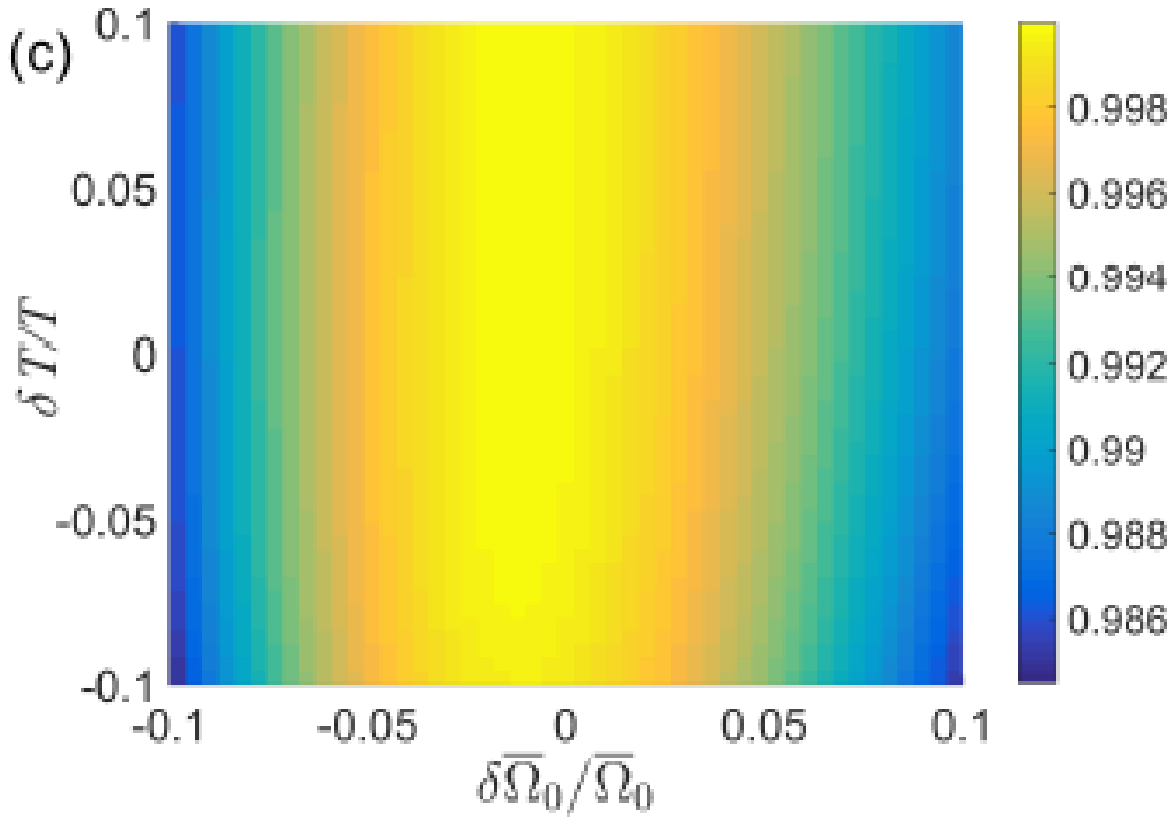}}
\caption
   {\label{fig8}
(a) The final fidelity $F(T)$ versus $\delta T /T$ and $\delta\lambda/\lambda$.
(b) The final fidelity $F(T)$ versus $\delta\overline{\Omega}_0 /\overline{\Omega}_0 $ and $\delta\lambda/\lambda$.
(c) The final fidelity $F(T$) versus $\delta T /T$ and $\delta\overline{\Omega}_0 /\overline{\Omega}_0$.
}
\end{figure}

In practice, the dissipations caused by decoherence usually place limits to the process, so it is helpful to check the performance of the protocol with different kinds of decoherence factors. In the present protocol, two main decoherence processes ought to be considered, i.e., cavity photon loss~(decay rate $\kappa$) and decays of the NV centers which are modeled by two characteristic decay rates $\gamma_{eg}$ between $|e\rangle$ and $|g\rangle$, $\gamma_{ef}$ between $|e\rangle$ and $|f\rangle$. Then the system is dominated by a master equation in the Lindblad form as following:
\begin{eqnarray}\label{eqc30}
  \dot{\rho}=i[\rho, H_{I}(t)]-\frac{\kappa}{2}(\hat{a}^{\dag}\hat{a}\rho-2\hat{a}\rho \hat{a}^{\dag}+\rho \hat{a}^{\dag}\hat{a})
  \cr
  -\sum_{p=1,2}\sum_{l=f,g}\frac{\gamma^{p}_{el}}{2}(\hat{S}^{p}_{el}\hat{S}^{p\dag}_{el}\rho -2\hat{S}^{p\dag}_{el}\rho \hat{S}^{p}_{el}+\rho \hat{S}^{p}_{el}\hat{S}^{p\dag}_{el}),
   \end{eqnarray}
where $\hat{a}$ is the annihilation operator of the cavity mode and $\hat{S}^{p}_{el}=|e\rangle_p\langle l|~(l=f,g)$.
For simplicity, we assume $\gamma^{p}_{el}=\gamma$.
In Fig.~7, we plot the final fidelity $F(T)$ versus $\kappa/\lambda$ and $\gamma/\lambda$. To make it clear, some of the data on Fig.~7 are shown in Table II. According to Fig.~7 and Table II, we can obtain: (i) $F(T)$ is robust to cavity decay since
the population of $|\psi_3\rangle$ approaches zero during the whole evolution, which can be seen from Fig.~5(b). (ii)  The spontaneous emissions of NV centers also have a pretty small detrimental effect on the protocol, i.e., $F(T)$ remains 0.9955 with $\gamma/\lambda=0.01,\kappa/\lambda=0$. This is easy to understand by seeing Fig.~5(b) since the populations of the intermediate states are well restrained with suitable parameters. In realistic experiments, the $Q$ factor of the WGM cavity can exceed $10^9$, which implies the decay rate is $\kappa\sim2\pi\times0.5$ MHz \cite{ParkNL0606,BarclayOE0917,SantoriNN1021,SpillanePRA0571}. The coupling strength between the NV center and the WGM can reach $\lambda/2\pi\sim0.3-1$ GHZ \cite{DujfNJP1012}. In addition, the decoherence rate of NV centers is about $\gamma/2\pi\sim13$ MHz \cite{SantoriPRL0697}.
Therefore, the above investigation about the choice of parameters is available in experiments.
 \begin{center}
{\bf Table II. Samples of the final fidelity $F(T)$ versus $\kappa/\lambda$ and $\gamma/\lambda$.}\\
{\small
  \begin{tabular}{cccc}
  \hline
  \hline
$ \kappa/\lambda $ \ \ \ \ \ \ \ \ \ \ \ \ \ \  & $\gamma/\lambda$ \ \ \ \ \ \ \ \ \ \ \ \ \ \  &$F(T)$ \\
\hline
$0.01\ \ \ \ \ \ \ \ \ \ \ \ \ \ $ & $0\ \ \ \ \ \ \ \ \ \ \ \ \ \ $ & $0.9965$\\
$0\ \ \ \ \ \ \ \ \ \ \ \ \ \ $ & $0.01\ \ \ \ \ \ \ \ \ \ \ \ \ \ $ & $0.9955$\\
$0.02\ \ \ \ \ \ \ \ \ \ \ \ \ \ $ & $0.02\ \ \ \ \ \ \ \ \ \ \ \ \ \ $ & $0.9848$\\
$0.03\ \ \ \ \ \ \ \ \ \ \ \ \ \ $ & $0.03\ \ \ \ \ \ \ \ \ \ \ \ \ \ $ & $0.9783$\\
$0.04\ \ \ \ \ \ \ \ \ \ \ \ \ \ $ & $0.04\ \ \ \ \ \ \ \ \ \ \ \ \ \ $ & $0.9713$\\
\hline
 \hline
\end{tabular}
}
\end{center}

Furthermore, to check the stability, we also investigate the influence of the experimental imperfect operations. Here the main imperfect variations are on $T$, $\overline{\Omega}_0$ and $\lambda$, which are denoted by $\delta T$, $\delta\overline{\Omega}_0,$ and $\delta\lambda$, respectively. The final fidelity $F(T)$ of state $|\psi_5\rangle$ versus (a) $\delta T/T$ and $\delta\lambda/\lambda$, (b) $\delta\overline{\Omega}_0 /\overline{\Omega}_0 $ and $\delta\lambda/\lambda$, (c) $\delta T /T$ and $\delta\overline{\Omega}_0 /\overline{\Omega}_0$ are plotted in Fig.~8, respectively.
According to Fig.~8, we can obtain the following results: (i) Seen from Figs.~8(a) and 8(b), $F(T)$ is insensitive to the variation $\delta\lambda$. The reason is obvious, as we have shown in Fig.~5(a), $F(T)$ is nearly 1 when $\lambda\geq 20/T$. While in the protocol, we select a suitable coupling constant $\lambda=30/T$. (ii) According to Figs.~8(a) and 8(c), the final fidelity $F(T^{\prime})$ is robust against the variation $\delta T$. The final fidelity is almost unchanged when both $\delta T/T,\delta\lambda/\lambda\leq 10\%$. (iii) Variation $\delta\overline{\Omega}_0$ influences the final fidelity mainly. Even so, when $\delta\overline{\Omega}_0/\overline{\Omega}_0=10\%$, the final fidelity is still higher than 0.98. This indicates that the present protocol holds robust against the variation $\delta\overline{\Omega}_0$ as well. Therefore, it can be concluded that the robustness against the imperfect operations and the decoherence are all quite nice for the present protocol.

\section{CONCLUSION}\label{section:IV}

We have worked out a framework to inverse engineer a Hamiltonian and construct shortcuts to the adiabatic passage by using picture transformation in SU(2) symmetric systems.
The constructed protocol holds the characteristics of flexibility since it frees the choice of parameters. Moreover, it is shown that the present protocol contains the results of the LR-based invariant method.
To show the effectiveness of the protocol and make a demonstration, we apply the protocol to inversely design a Hamiltonian to execute population transfer inversion between two distant NV centers. With the protocol, the maximal populations of the lossy intermediate states and the amplitudes of pulses can be controlled by two corresponding control parameters. Numerical simulation demonstrates that the protocol is fast and robust against
the decoherence and operational imperfection.
Note that, besides exploring robustness by numerical solution, there are other methods can be utilized to investigate and improve the robustness of the protocol. For example,
 in Ref.~\cite{Daems13111}, the authors reduced the robustness issue to nullifying integrals that cancel out the derivatives of the excitation profile order by order and  rendered an explicit analytic solution to tackle the robustness issue.
    In Ref.~\cite{Ruschhaupt1214}, the stability of the shortcuts to adiabaticity was investigated by examining the amplitude-noise error and systematic errors.
 Both of these methods provide us an alternative perspective to investigate the robustness issue.

Also, we hope the present protocol might be useful for implementing
fast and noise-resistant quantum information processing in multi-qubit systems
within current technology.
Of course, we should also address the fact that for a more complicated system, more shortcut methods deserve further exploration.

\section*{ACKNOWLEDGEMENT}

 This work was supported by the National Natural Science Foundation of China under Grants No. 11575045 and No. 11374054, the Major State Basic Research
Development Program of China under Grant No. 2012CB921601, and the Natural Science Foundation of Fujian Province under Grant No. JAT160081.

\newpage

\end{document}